\documentclass{article}

\usepackage{amssymb}
\usepackage{amsmath}
\usepackage{graphicx}
\DeclareGraphicsExtensions{.jpg,.png}

\usepackage[export]{adjustbox}

\usepackage{color}

\begin{document}

\title{Uncertainty quantification in covid-19 spread: lockdown effects}

\author{Ana Carpio\footnote{Departamento de Matem\'atica Aplicada, Universidad Complutense, 28040 Madrid,  Spain, ana\_carpio@mat.ucm.es},
Emile Pierret \footnote{CMLA, ENS Paris-Saclay, 91190 Gif-sur-Yvette, France}}

\maketitle

{\bf Abstract.}
We develop a Bayesian inference framework to quantify uncertainties in epidemiological 
models. We use SEIJR and SIJR models involving populations of susceptible, exposed, 
infective, diagnosed, dead and recovered individuals to infer from covid-19 data rate constants, as well as their variations in response to lockdown measures.  To account for confinement, we distinguish two susceptible populations at different risk: confined and unconfined. We show that transmission and recovery rates within them vary in response to facts. 
A key unknown to predict the evolution of the epidemic is the fraction of the population affected by the virus, including asymptomatic subjects. Our study tracks its time evolution with quantified uncertainty from available official data, limited, however, by the data quality. We exemplify the technique with data from Spain, country in which late drastic lockdowns were enforced for months. In late actions and in the absence of other measures, spread is delayed but not stopped unless a large enough fraction of the population is confined until the asymptomatic population is depleted. To some extent, confinement could be replaced by strong distancing through masks in adequate circumstances.

\section{Introduction}
\label{sec:intro}

Since the outbreak of the current covid-19 pandemic \cite{med2,med1}, 
Health  Services worldwide report daily data about the status of the epidemic,
which serve as a guide for the design of non-pharmaceutical interventions \cite{policy,reproduction}. An increasing number of mathematical 
studies  assess the efficacy of different policies 
\cite{lockdown_india,spatial,bayesian_hierarchical,bayesian_germany,
bayesian_law,reproduction}.
Moreover, mathematical models and data analysis are employed to estimate 
relevant epidemiological parameters 
\cite{sars_diff,reproduction,covid1,covid2,serial_data,adjoint_assimilation}
and to try to forecast the evolution
\cite{fuzzy,sequential_assimilation,alessandra,susceptible1,
seir_in,susceptible2,machine}. While some of this research is based on
direct data analysis \cite{reproduction,serial_data}, machine learning
techniques \cite{fuzzy,machine} or empirical laws for different populations
\cite{bayesian_law}, the use of balance equations
to predict  population dynamics is a common approach.

After the pioneering work of Kermack and McKendrick \cite{sir}, SIR type 
models have become a standard tool in epidemiological studies  \cite{book}. 
The specific structure of the selected models depends on the available
information and on assumptions about the epidemic spread \cite{models}.
Basic SIR models involve populations of susceptible $S$, infected $I$,
and recovered $R$ individuals, expecting immunity of the latter \cite{sii,bayesian_germany,alessandra,adjoint_assimilation}. 
SEIR variants distinguish also the individuals exposed to the virus $E$, which 
may become infective \cite{covid1,seir_in}. Immunity of the recovered is 
suppressed in SEIRS systems \cite{susceptible1,susceptible2}.
During the  2002-04 SARS (Severe Acute Respiratory Syndrome) outbreak, 
these models were adapted to describe the SARS epidemic in different 
countries by singling out the diagnosed infective $J$ \cite{sars_first, sars_diff}, 
becoming SEIJR or SIJR models. Diagnosed individuals are isolated.
The virus SARS-CoV-2 responsible for the illness covid-19 belongs to the 
same family as the virus SARS-CoV, responsible for SARS. The epidemics
triggered by them share some features, such as the role of asymptomatic
individuals in superspread events, see \cite{covid2} for a quantification of the 
fraction of asymptomatic population during covid-19 spread following this 
approach.  
Here, we will study the effect of confinement measures on covid-19 spread by
distinguishing two susceptible SEIJR populations:  confined and non confined.

To have a predictive value, we must fit the model parameters to
available data. This can be done applying optimization  or adjoint-based data 
assimilation techniques to reduce  the difference between recorded data and model 
predictions for selected parameters \cite{sars_diff,adjoint_assimilation}, for 
instance. However, data for epidemiological studies are subject to many sources 
of noise and uncertainty. In the case of the current covid-19 pandemic, different 
countries, and regions within them, define the diagnosed, recovered and dead 
individuals they count in their official reports in different ways. The number of dead 
individuals may refer only to patients who die in hospitals or include also deaths
at homes and care homes. Furthermore, the death of covid patients with previous
health issues may be officially attributed to other causes. On the other hand,
the number of diagnosed individuals may refer only to cases confirmed  by a 
PCR (Polymerase chain reaction) test or include also positive antibody tests,
or probable cases with  compatible symptoms and  clinical history. 
Moreover, the results of tests may arrive with a variable delay, which results 
in fluctuations and exclusions. Undated cases may not be counted 
at all. Tests repeated for the same individuals may be counted as different.
Additionally, the number of tests performed varies largely over the weeks due to 
supply shortages and changes in local testing policies, and the accuracy of 
the tests employed may fluctuate, yielding false negatives or positives. 

\begin{figure}[h!]
\centering
(a) \hskip 6cm (b) \\
\includegraphics[width=7cm,valign=t]{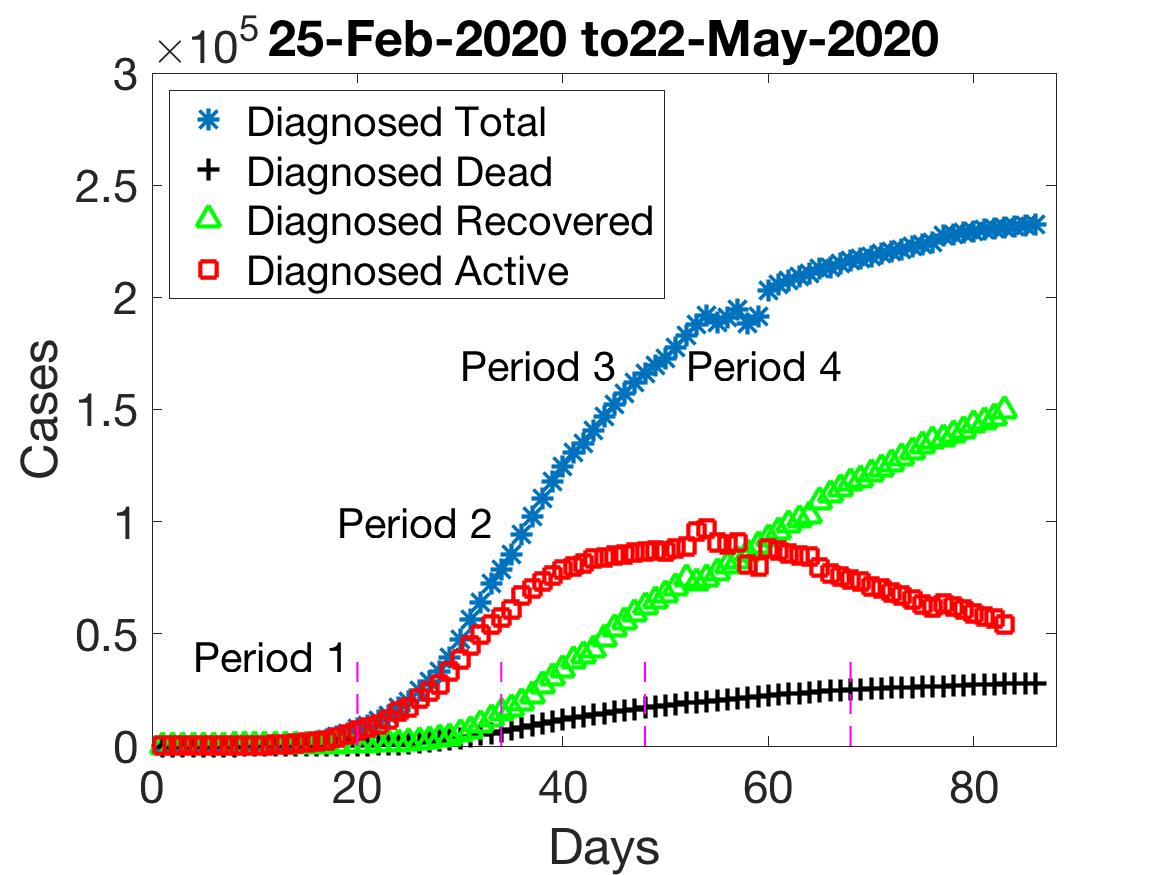} \hskip -5mm
\includegraphics[width=7cm,valign=t]{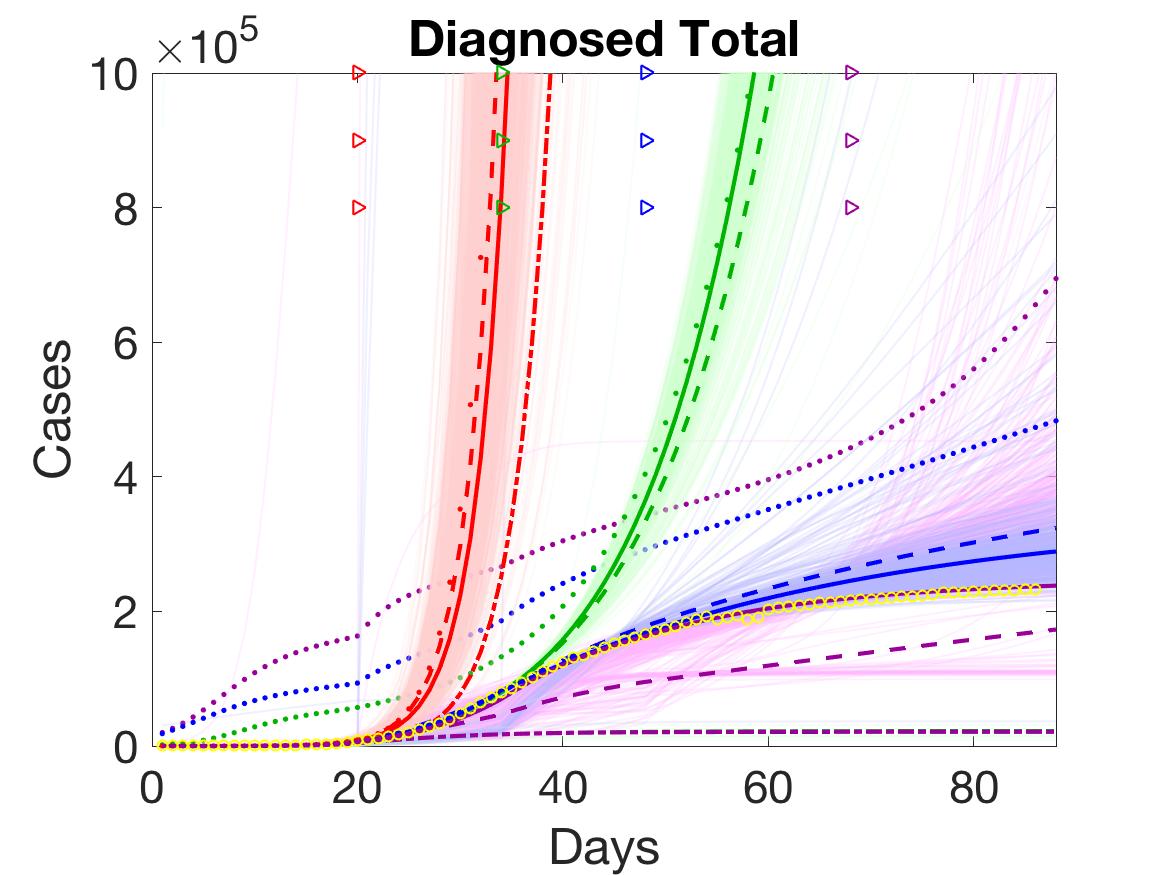}
\caption{(a) Daily counts of diagnosed, recovered and dead individuals
(PCR confirmed) in Spain since February 25th, 2020, until May 22th, 2020 
\cite{dataspain}. 
After an initial period of uncontrolled spread (Period 1), borders were closed,
while all the population being  able to work online, or not working in basic activities, 
was confined at home in the whole country (Period 2): education, administration, 
tourism, shopping, leisure activities... Lockdown was later extended to all non 
essential activities (Period 3). Only food and medical supplies, healthcare, security, 
essential transport and essential production remained active. Confinement was 
then released by stages, first some workers (Period 4), then the rest, while introducing recommendations for the use of masks and social distancing.
(b) SEIJR based Bayesian inference and predictions for the total number 
of diagnosed individuals using counts  from Period 1 (red),
Periods 1-2 (green), Periods 1-2-3 (blue) and Periods 1-2-3-4 (magenta). For each
of them, top coloured triangles separate the inference from the prediction part of the 
simulations. True data are marked by yellow circles. Solid curves correspond to best fits,
dashed curves and dotted curves to different types of sample averages. Shaded areas
and dotted curves define uncertainty regions, see Section \ref{sec:contention} for a 
discussion.}
\label{fig0}
\end{figure}

Uncertainty in the data propagates to any predictions based on them.
Instead of fixing specific guesses for the model coefficients, it is
convenient to explore approaches that quantify uncertainty \cite{bayesian_hierarchical,capistran,bayesian_germany,bayesian_law,covid2}.
Unlike most work which does not distinguish undocumented and 
documented infected individuals, here we follow the SEIJR approach and 
compare data to model predictions of diagnosed infected $J$
\cite{sars_first,sars_diff,covid2}, including quarantine measures for them
and taking into account the diagnose rate due to testing.
We develop a general framework to infer SEIJR model coefficients from
data  with quantified uncertainty, taking into account confinement measures 
as they are sequentially enforced or lifted by means of two populations: 
confined and unconfined. 
This allows us to analyze variations in the  model rates and in the 
distributions of the different populations a time grows as a result of the 
measures implemented, including undiagnosed infected individuals and 
asymptomatic individuals. 
We focus on the case of Spain, where  drastic late global lockdowns  were 
enforced at the same time in the whole country, producing well 
differentiated periods in the data along a long time period, see Fig. \ref{fig0}.
The situation is quite different from  the german case, in which mild measures
were implemented very early to curb the spread \cite{bayesian_germany}, 
the italian case, where strongth spatiotemporal differences between regions 
occurred \cite{spatial,alessandra}, and  from studies of initial stages  
 \cite{covid1,seir_in}.  Nevertheless, our methods apply to data for diagnosed, 
dead and recovered individuals from any other country. 
The key idea is introducing a susceptible subpopulation at lower risk, 
which might also be achieved by milder measures such as generalized 
distancing through masks instead of confinement in a closed system,
no individuals enter or exit the system.
The analysis of migration and spatial dynamics are relevant topics \cite{sequential_assimilation,spatial,covid2}, still out of the scope of the  
present study.


The next sections are organized as follows. Section \ref{sec:sijr} recalls the
structure of SEIJR models. We intend to quantify uncertainty when fitting 
these models to data from the current covid-19 pandemic. 
We use the SIJR simplification for the initial stage of the outbreak, before 
contention measures were taken, and compare to the full SEIJR results. 
SIJR predictions usually underestimate the total number of affected people.
Section \ref{sec:fitting} explains how to obtain guesses of model parameters, 
which  play the role of prior knowledge for the Bayesian studies 
in Section \ref{sec:bayesian}. Section \ref{sec:free}  analyzes the initial stage 
while Section \ref{sec:contention} considers the effect of contention measures,
with spanish data.  
We adapt the SEIJR framework to study parameter uncertainty 
through the different stages, inferring also key magnitudes such as the time 
evolution of the number of asymptomatic and undiagnosed individuals affected 
by the virus,  or the global number of affected people. In  late interventions, 
and in the absence of other preventive measures, spread is delayed but not 
stopped unless a large enough fraction of the population is confined for a long 
enough time, until the number of asymptomatic and undiagnosed individuals is 
depleted.
Once confinement is over, the usage of masks plays a similar role keeping a 
fraction of the population at a lower risk in a closed system.
Section \ref{sec:conclusions} summarizes our conclusions.

\section{SEIJR models for SARS and Covid-19 type epidemics}
\label{sec:sijr}

SEIJR models involving populations of susceptible (S), exposed (E), infective (I),  
diagnosed (J), and recovered (R) individuals were proposed in \cite{sars_first}
to study  the spread of the 2002-04 SARS outbreak. Here, we will adapt them
to describe contention measures for covid-19.
Considering  two populations  $S_1$ and $S_2$ of different susceptibility,
the model takes the form:
\begin{eqnarray} \begin{array}{c}
\displaystyle {d S_1 \over dt} = -\beta S_1(t) { I(t)+qE(t)+\ell J(t) \over N}, \\[1.5ex]
\displaystyle {d S_2 \over dt} = -\beta p S_2(t) { I(t)+qE(t)+\ell J(t) \over N}, \\[1.5ex]
\displaystyle {d E \over dt} = \beta (S_1(t) + pS_2(t)) {I(t)+qE(t)+\ell J(t) \over N} 
- k E(t), \\[1.5ex]
\displaystyle {dI \over dt} = kE(t) - (\alpha+\gamma_1+\delta) I(t), \\[1.5ex]
\displaystyle {dJ \over dt} = \alpha I(t) - (\gamma_2+\delta) J(t), \\[1.5ex]
\displaystyle {dR \over dt} = \gamma_1 I(t) + \gamma_2 J(t), \\[1.5ex]
\displaystyle{dD \over dt} = \delta I(t) + \delta J(t),
\end{array} \label{SEIJR} 
\end{eqnarray}
where $N=S_1+S_2+E+I+J+R+D$ is the total population number, which remains
constant. $D$ is the number of dead individuals.
The exposed $E$ are a class of asymptomatic and possibly infectious individuals.
The possibility of transmission from exposed individuals $E$ is represented 
by the parameter $q$. They may progress to the infective state $I$ at a rate $k$.
The class $I$ is composed of symptomatic, infectious, and undiagnosed individuals.
Infectious individuals $I$ become diagnosed $J$ at a rate $\alpha$.
The recovery rate of the infective $I$ is $\gamma_1$, whereas the recovery rate 
of  the diagnosed $J$ is $\gamma_2$. The recovered individuals $R$ keep track
of the cumulative number of sick individuals who become healthy again.
Diagnosed individuals $J$ are isolated from the rest. Their reduced impact on
transmission is represented through a parameter $\ell$. Mortality of infected
$I$ and diagnosed $J$ individuals caused by the virus is denoted by $\delta$.
Finally, $\beta$ represents the transmission rate: how susceptible $S$ individuals
become virus spreaders. Time is measured in days.

The model has to be complemented with initial conditions. This fact introduces an
additional parameter $t_{\rm in}$ to locate the time at which local spread started
\cite{sars_diff}. Other approaches assume the initial data unknown instead
\cite{bayesian_germany}, in our case that choice would increase considerably 
the number of unknowns.
Furthermore, we consider that the risk of infection for $S_2$ is lower 
than the risk for $S_1$ by a factor $p$. The total population is partitioned 
as $S_1= (1-\rho) S$, $S_2= \rho S$, $\rho$ being the fraction of the 
susceptible population $S$ at a lower risk of infection. 
The risk might vary due to specific characteristics 
of the population (age, sex, genes) \cite{sars_first}. Here, variations will be due to 
confinement/protection measures enforced on part of the population.

\begin{table}[h!]
\small \centering
\begin{tabular}{|c|c|c|}
  \hline
  Par. & Definition & Guess  \\
  \hline
  $\beta$ & Transmission rate per day & \\
  $k $ & Rate of progression to the infectious state per day &   \\ 
  $\alpha$ & Rate of progression from infective to diagnosed per day  & 1/5-1/6 (stats) \\
  $\gamma_1$ & Rate at which infectious individuals recover per day &
  $\gamma_1^{-1} = \gamma_2^{-1} + \alpha^{-1}$ \\
  $\gamma_2$ & Rate at which diagnosed individuals recover per day & 1/10-1/11
  (stats) \\
  $\delta$ & covid-19 induced mortality per day & 1/10-1/11 (stats) \\ 
  $\ell$ & Relative measure of isolation of diagnosed cases & 1/14 (practice) \\ 
  $q$ & Relative measure of infectiousness for the exposed & \\
  $p$ & Reduction in risk of covid-19 infection for class $S_2$ & \\
  $t_{\rm in}$ & Time at which local spread starts & \\
  $\rho $ & Fraction of the population at a lower risk & \\ 
  \hline
\end{tabular}
\caption{SEIJR model parameters. Guesses from clinical observation when
available \cite{renave}.}
\label{table1}
\end{table}

Two constraints are usually imposed on the parameters: 1) $\alpha > \gamma_1$ and
2) $\gamma_2^{-1} = \gamma_1^{-1} - \alpha^{-1}$  \cite{sars_first}. Moreover,
the following  expression for the basic reproduction number \cite{sars_first} holds 
\begin{eqnarray*}
{\cal R}_0 = \beta (\rho + p(1-\rho)) \left(
{q\over k} + {1\over \alpha +\gamma_1+ \delta}
+ {\alpha \ell \over (\alpha +\gamma_1+ \delta)(\gamma_2+ \delta)
}\right).
\end{eqnarray*}
The reproduction number represents the expected number of cases  immediately 
originated by one case in a population where all individuals are susceptible to infection, 
that is, no other individuals are infected or immunized (naturally or through vaccination).
Instead, the effective reproduction number ${\cal R}_e$ is just the number of cases 
produced in the current state of a population.

This type of models reproduces crudely some characteristics observed in SARS
epidemics, such as the emergence of symptomatic and asymptomatic individuals, 
superspread events and unequal susceptibility, for instance. 
We will use them here with data from the current covid-19 epidemic.  First guesses 
for some of the model parameters can be estimated from average observations, see 
Table I. First guesses for two key parameters, $t_{\rm in}$ and $\beta$ can be
obtained from simplified SIJR approximations, as we explain in the next section.

\section{Fitting the initial stages of the outbreak}
\label{sec:fitting}

The SEIJR models  we have introduced assume that 
1) spread takes place in a closed system, 
2) the death rate is the same for everybody (death by other causes is neglected),
3) the recovered have immunity,  
4) the diagnosed are isolated, and
5) time delays in responses are neglected.
Assuming further that: 
6) the exposed phase $E$ is neglected, 
7) the susceptibility degree is not distinguished $S_1=S_2=S$, $p=1$,
8) the infected are a small fraction of the whole population, so that
${S\over N} \sim 1$, we obtain a SIJR simplification \cite{sars_diff}:
\begin{eqnarray}
{d S \over dt} = - \beta (I + \ell J), \label{ecS} \\
{d I \over dt} = (\beta - (\alpha + \gamma_1 + \delta) ) I + \ell \beta J,  
\label{ecI} \\
{d J \over dt} = \alpha I - (\gamma_2 + \delta) J,  \label{ecJ} \\
{d R \over dt} = \gamma_1 I + \gamma_2 J,  \label{ecR} \\
{d D \over dt} = \delta (I + J), \label{ecD} \\[1ex]
N= S + I + J + R +D,  \label{ecN} \\[1ex]
S(t_{\rm in}) = N-1, \, I(t_{\rm in}) =1,  \, J(t_{\rm in})=R(t_{\rm in})=0=D(t_{\rm in}). 
\label{ic}
\end{eqnarray}
Here, $N=S+I+J+R+D$ is the total population number, which remains
constant. SIJR models allow us to fit important parameters, such as the 
transmission rate $\beta$ and the onset of local spread $t_{\rm in}$,
which determine the exponential growth in the initial stages. Their solutions
admit analytic expressions, detailed in Appendix \ref{sec:explicit}.
Thanks to that fact,  they have been used to analyze the influence of isolation 
measures on the inflexion point, see  \cite{sars_diff} and references therein. 
Notice that sign balances in (\ref{ecI}) govern the increase of the number 
of infected people.

In the SIJR model (\ref{ecS})-(\ref{ic}), we have to fit the 
parameters $\alpha,$ $\gamma_1,$ $\gamma_2,$ $\delta,$ $\ell,$ $\beta$, 
as well as $t_{\rm in}$, defined as the time at which $I(t_{\rm in})=1$.  This can be 
done starting from educated guesses and optimizing a cost functional 
with respect to them. The clinical information collected during the current
pandemic \cite{renave} yields tentative average values for the rates
$\alpha,$ $\gamma_1,$ $\gamma_2,$ $\delta,$ and for $\ell$, collected
in Table \ref{table1}.  We then seek to fit the remaining parameters by optimizing
a cost. A popular choice is
\begin{eqnarray}
f(\beta,t_{\rm in})= {1\over 2}\sum_{j=1}^L (\tilde J(\beta,j+t_{\rm in}) -\tilde y_j)^2,
\label{cost1}
\end{eqnarray}
where $\tilde y_j$, $j=1,...,L$, are cumulative numbers of diagnosed people for 
$L$ days and the cumulative variable $\tilde J$ solves $\tilde J' = \alpha I$, 
$\tilde J(0)=0,$ with $I$ given by (\ref{ecI}). This variable $\tilde J$ is in fact the total 
cumulative number of diagnosed individuals, obtained adding to $J$ the diagnosed 
recovered $R_{J}$ and the diagnosed dead $D_{J}$, solutions of
\begin{eqnarray}
R_J' =  \gamma_2 J, \quad  D_J' =  \delta J,  \quad
R_J(t_{\rm in}) = D_J(t_{\rm in}) = 0. \label{RjDj}
\end{eqnarray}
This is an important distinction.
Note that equation (\ref{ecJ}) discounts the diagnosed people who recover or
die, thus $J$ tracks only the active diagnosed cases. In practice, only the
the diagnosed recovered $R_{J}$,  the diagnosed dead $D_{J}$ and the
diagnosed active $J$ or total $\tilde J$ are recorded by Health Care Systems, 
since the contribution coming from undiagnosed infected cases is unknown.

SIJR models are particularly adequate for these fittings because solutions admit 
explicit expressions which reduce numerical errors when dealing with exponentially 
growing solutions, see Appendix 
\ref{sec:explicit}. We will  resort to the Levenberg-Marquardt-Fletcher algorithm 
\cite{lmf} to optimize the costs.

\begin{figure}[h!]
\centering
\includegraphics[width=6cm,valign=t]{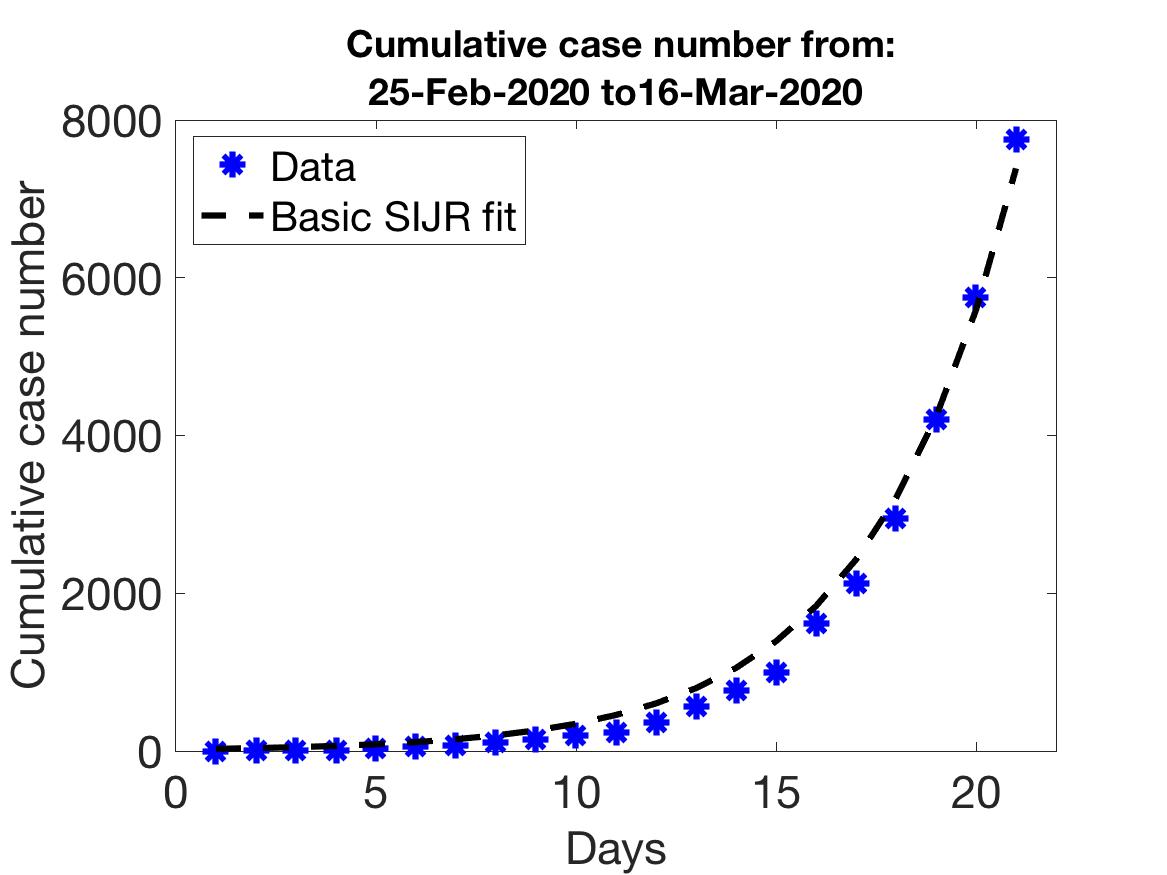}
\caption{Parameter guess for the first period of data in Fig \ref{fig0} (free
spread): $t_{\rm in}= 12.3388$, $\beta=0.6262$,
$\alpha=1/5$, $\gamma_1= 1/15$, 
$\gamma_2= 1/10$, $\delta= 1/10$, $\ell= 1/14$.}
\label{fig1}
\end{figure}

The final values we obtain for $t_{\rm in}$ and $\beta$ are $12$ days and $0.6262$, 
starting the optimization from initial guesses $10$ and $0.6$. This is consistent with 
the fact that deaths occurred as early as February 13 in Spain were proven to be caused 
by  covid-19. Notice that we are fitting a cumulative magnitude $\tilde J = J + R_{J} + 
D_{J}$.  Even if the fitting for $\tilde J$ is accurate, as Fig \ref{fig1} shows, the results 
worsen noticeably when we use these parameters to calculate $J$, $R_{J}$, and 
$D_{J}$ and compare with the data recorded for each of them.

We could improve the overall guess using these values as starting point
for an algorithm optimizing the cost  
\begin{eqnarray}
f(\alpha, \gamma_2,  \delta, \ell, \beta, t_{\rm in})=  {1\over 2}
\sum_{j=1}^L (\tilde J(\alpha, \gamma_2, \delta,  \ell, \beta, j+t_{\rm in}) 
- \tilde y_j)^2,
\label{cost2}
\end{eqnarray}
with respect to all of the parameters, or resorting to more detailed cost 
functionals.
However, our goal here is to quantify uncertainty in usual rough fits and 
predictions obtained with them. Therefore, we will use them as priors for 
the subsequent Bayesian studies.

\section{Uncertainty quantification by Bayesian techniques}
\label{sec:bayesian}

Bayes' theorem describes the probability of an event, based on prior knowledge about 
it \cite{somersalo}. According to it, the posterior probability of observing a finite
number of parameters  $\boldsymbol \nu$ given data $\mathbf d$ would be
\begin{eqnarray*}
p( \boldsymbol \nu | \mathbf d) = { p( \mathbf d | \boldsymbol \nu ) p(\boldsymbol \nu)
\over p(\mathbf d)}
\label{bayes}
\end{eqnarray*}
where $p( \mathbf d | \boldsymbol \nu )$ is a conditional probability (the likelihood of 
observing data $\mathbf d$ given parameters $\boldsymbol \nu$), and $p(\boldsymbol \nu)$ 
represents our prior knowledge on the parameters $\boldsymbol \nu$. The normalization
factor $p(\mathbf d)$ represents the probability of the data. It is also a marginal probability, 
which can be obtained integrating
$p( \mathbf d | \boldsymbol \nu ) p(\boldsymbol \nu)$ with respect to $\boldsymbol \nu.$

Let us fit our problem in this framework. The parameters are the model parameters, that is,
\begin{eqnarray}
\boldsymbol \nu= (t_{\rm in}, \beta,  \gamma_{2}, \delta, \alpha,  \ell), \quad
\gamma_{1}^{-1} =  \gamma_{2}^{-1}  + \alpha^{-1},
\label{nu}
\end{eqnarray}
for the  SIJR  model or, for SEIJR,
\begin{eqnarray}
\boldsymbol \nu= (t_{\rm in}, \beta,   \gamma_{2}, \delta, \alpha,  \ell, q,
p, k), \quad \gamma_{1}^{-1} =  \gamma_{2}^{-1}  + \alpha^{-1}.
\label{nu2}
\end{eqnarray}
Then, the prior distribution, the likelihood and the posterior distribution are defined as
follows.

\subsection{Prior distribution}
\label{sec:prior}

For the prior distribution, we use a parameter guess  $\boldsymbol \nu_0$ as 
the mean of a  multivaluate normal distribution with a covariance matrix 
$\mathbf G_{\rm pr}$ constructed from the deviations of each  variable
\begin{eqnarray} \begin{array}{l}
p(\boldsymbol \nu)  
={1\over (2 \pi)^{n/2}}  {1\over \sqrt{|\mathbf G_{\rm pr} |}}
\exp(-{1\over 2}
(\boldsymbol \nu - \boldsymbol \nu_0)^t \mathbf G_{\rm pr}^{-1} 
(\boldsymbol \nu - \boldsymbol \nu_0) ),
\end{array} \label{prior}
\end{eqnarray}
where $n$ is the number of parameters.  We choose a diagonal covariance matrix 
$\mathbf G_{\rm pr}$ with elements $\sigma_i^2$, $i=1,...,n$. In practice, we have 
to modify this proposal because our parameters are always positive and gaussians 
may produce negative values. Thus, we set
\begin{eqnarray}  
p_{\rm pr}(\boldsymbol \nu)  
= \left\{ \begin{array}{cc} \exp(-{1\over 2}
(\boldsymbol \nu - \boldsymbol \nu_0)^t \mathbf G_{\rm pr}^{-1} 
(\boldsymbol \nu - \boldsymbol \nu_0) ), & \nu_j \geq 0, j=1,...,n, \\
0, & \nu_j <0, \mbox{ for some $j$.}
\end{array} \right. \label{prior}
\end{eqnarray}
This will be our choice of prior distribution $p_{\rm pr}( \boldsymbol \nu)$.
We do not need to calculate the normalization factor for later use, since
our sampling techniques do not require it.

\subsection{Likelihood}
\label{sec:like}

For the conditional probability density $p( \mathbf d | \boldsymbol \nu)$ 
we set
\begin{eqnarray} \label{likelihood}
p( \mathbf d  | \boldsymbol \nu) = {1 \over (2\pi)^{L/2} \sqrt{|\mathbf G_{\rm n}|}} 
\exp \Big(- {1 \over 2} \| \mathbf f( \boldsymbol \nu) - \mathbf d  
 \|^2_{\mathbf G_{\rm n}^{-1}} \Big),
\label{likelihood}
\end{eqnarray}
where $\| \mathbf v \|_{\mathbf G_{\rm n}^{-1}}^2 = \mathbf {\overline v}^t 
\mathbf G_{\rm n}^{-1} \mathbf v$, 
$\mathbf G_{\rm n}$being the covariance matrix representing the noise in the 
data $\mathbf d$, and $\mathbf f( \boldsymbol \nu)$   the observation
operator. We assume additive Gaussian noise, i.e., the 
observations and true parameters would be related by
\begin{equation} \label{noisydata}
  \mathbf d = \mathbf f(\boldsymbol \nu_{true}) + \boldsymbol \varepsilon.
\end{equation}
Here, the noise $\boldsymbol \varepsilon$ is distributed as a
multivariate Gaussian ${\cal N}(0,\mathbf G_{\rm n})$ with mean zero and 
covariance matrix $\mathbf G_{\rm n}$. 

In practice, the data available are daily cumulative counts of diagnosed individuals 
$j_m$, diagnosed recovered $r_m$ and diagnosed dead $d_m$,
$m=1,...,M$, see \cite{dataspain}. Putting the three blocks of data together we have
\begin{eqnarray}
\mathbf d = (\tilde j_1,...,\tilde j_M, r_1,...,r_M, d_1,..., d_M),
\label{data}
\end{eqnarray}
where $\tilde j_m= j_m - r_m - d_m$ are the active diagnosed, those who are neither 
dead nor recovered. Following \cite{pierret}, we define the observation operator as
\begin{eqnarray}
\mathbf f(\boldsymbol \nu) = (J(1), ..., J(M), R_J(1), ..., R_J(M), D_J(1), ..., D_J(M)),
\label{measurement}
\end{eqnarray}
where the dynamics of the diagnosed recovered $R_J$ and diagnosed dead $D_J$ are 
governed by (\ref{RjDj}) whereas the diagnosed individuals $J$ in which the infection is 
active are governed by (\ref{ecJ}) for SIJR (see Appendix \ref{sec:explicit} for analytic 
expressions) or (\ref{SEIJR}) for SEIJR. In (\ref{likelihood}),
we compare these observations to the data $\boldsymbol d$ using the distance
$ {1 \over 2} \| \mathbf f( \boldsymbol \nu) - \mathbf d  \|^2_{\mathbf G_{\rm n}^{-1}}$.
To simplify, we consider the noise level 
for all observations to be uncorrelated, so that $\mathbf G_{\rm n}$ is a real 
diagonal matrix,   $\mathbf G_{\rm n}=  {\rm diag}(\sigma_1^2,\ldots,\sigma_L^2)$, 
and set all the variances for the same magnitude equal to a constant 
$\sigma_{\rm J}^2,$ $\sigma_{\rm R}^2,$$\sigma_{\rm D}^2.$ Thus, 
$\sqrt{|\mathbf G_{\rm n}|}=\sigma_{\rm J}^M \sigma_{\rm R}^M
\sigma_{\rm R}^M$, where $L=3M$ is the number of data considered.
Note that these cost functionals require more information than those based
on total case counts: we distinguish diagnosed individuals who are dead, 
recovered and still sick, and  compare with model predictions for them discarding
the contribution of the undiagnosed, unlike \cite{bayesian_germany,sars_diff}.

\subsection{Posterior distribution}
\label{sec:posterior}

Combining  (\ref{prior}) with (\ref{likelihood}) and neglecting normalization constants, 
the posterior density becomes, up to multiplicative constants,
\begin{eqnarray}
p_{\rm pt}(\boldsymbol \nu) \sim \exp \left( -{1\over 2} \|
\mathbf f( \boldsymbol \nu) - \mathbf d  \|^2_{\mathbf G_{\rm n}^{-1}} -
 {1\over 2} \|  \boldsymbol \nu - \boldsymbol \nu_0 \|_{\mathbf G_{\rm pr}^{-1}}^2
 \right). \label{posterior}
\end{eqnarray}

By sampling this posterior distribution, we can visualize the uncertainty in the 
inference of parameters for a given data set. To do so, we will resort to Markov 
Chain Monte Carlo Sampling  \cite{sergei,hammer}. Once we 
have a large collection of samples, we can extract information from the model 
(\ref{ecS})-(\ref{ecN}) with quantified uncertainty, such as the global number of  
people who have been affected by the virus the last day of the period we are 
considering. In the next sections  we  exemplify the procedure for 
the different stages of the epidemic as observed in Figure \ref{fig0}.

\section{Uncertainty in the initial stage}
\label{sec:free}

\begin{figure}[h!]
\centering
(a) \hskip 3.5cm (b) \hskip 3.5cm (c) \\
\includegraphics[width=4.8cm]{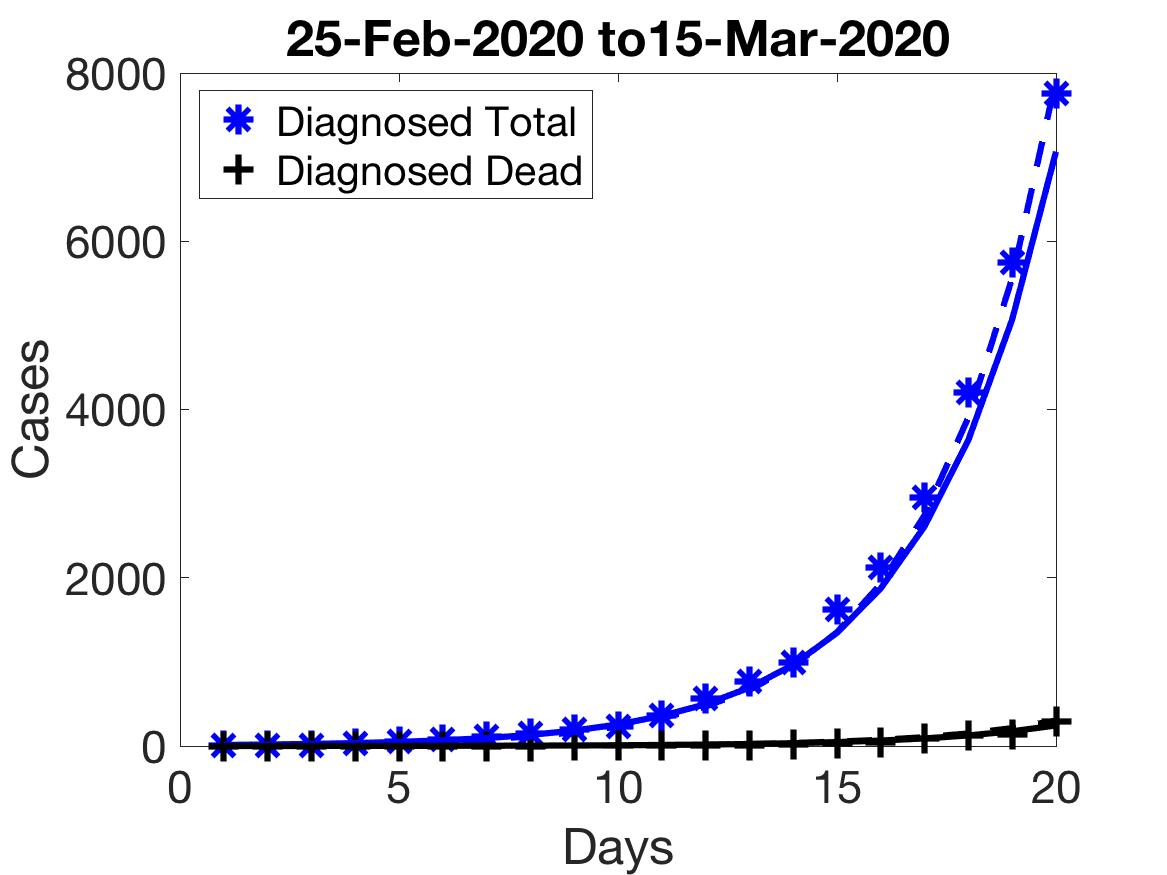} \hskip -5mm
\includegraphics[width=4.8cm]{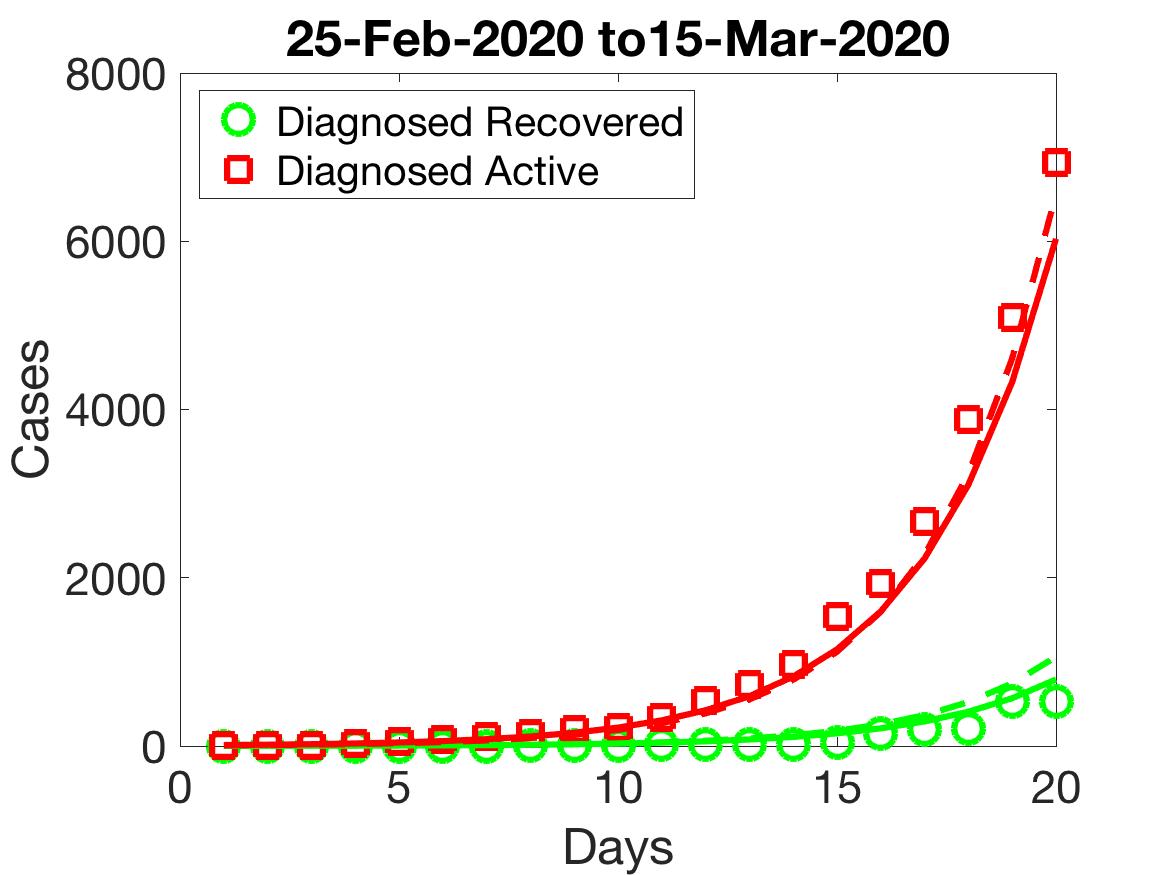} \hskip -5mm
\includegraphics[width=4.8cm]{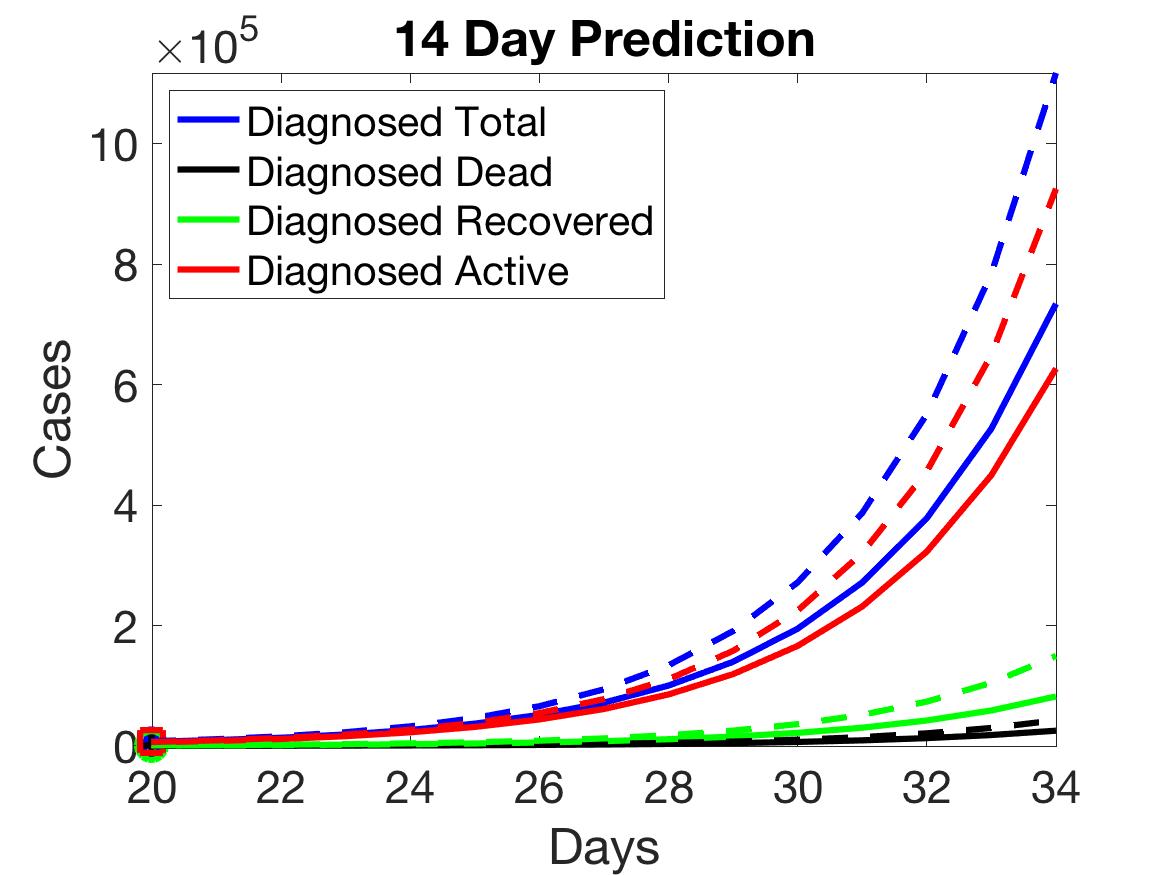} \\
(d) \hskip 6cm (e)  \\
\includegraphics[width=6.8cm]{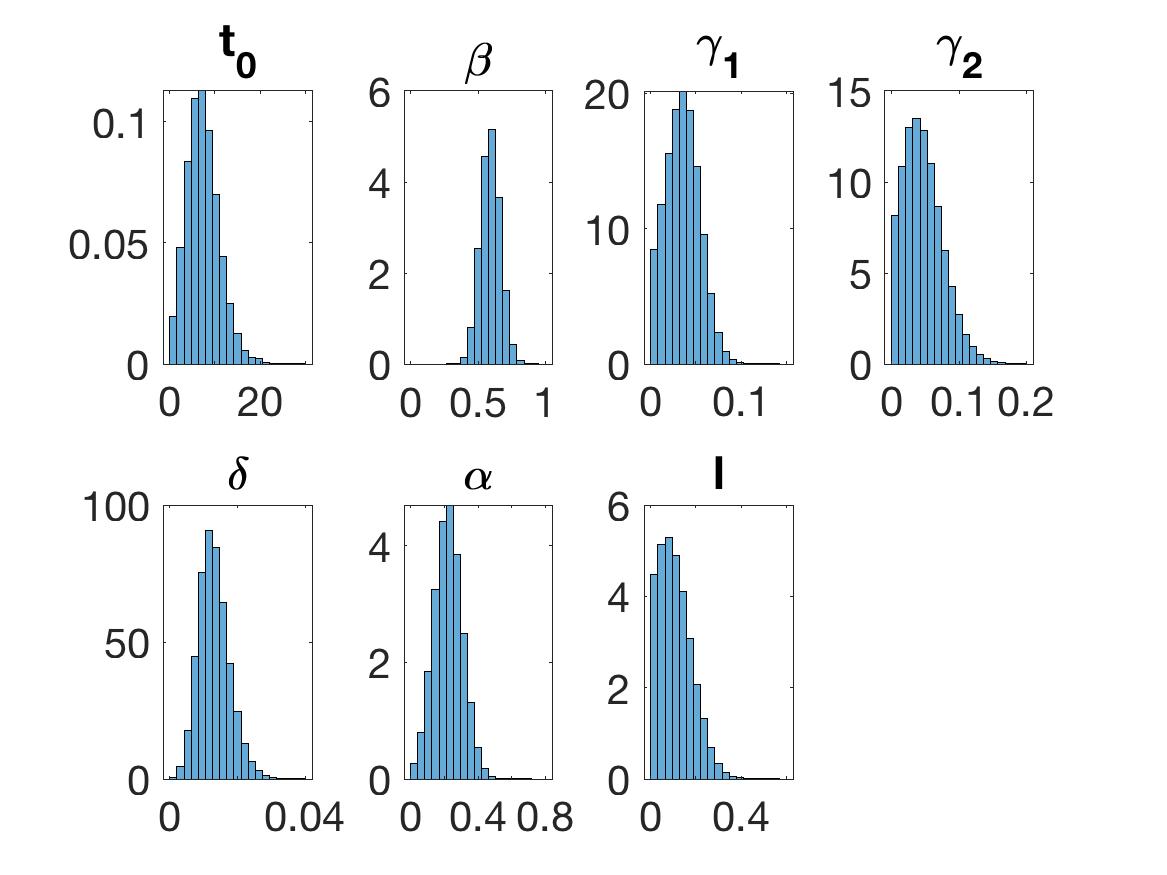} \hskip -4mm
\includegraphics[width=6.8cm]{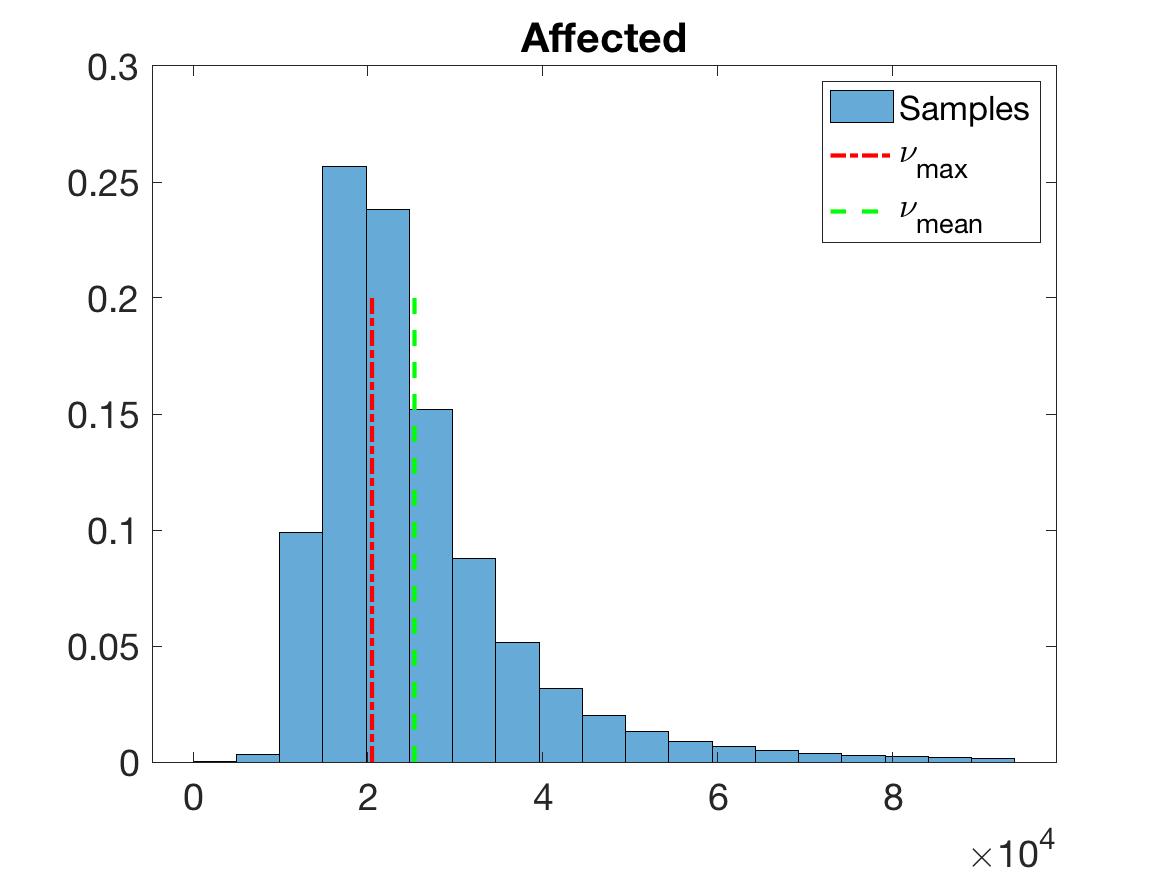} 
\caption{Initial stage (free spread): 
(a) counts of diagnosed and dead cases
compared to SIJR solutions of (\ref{ecJ}) and (\ref{RjDj}) for 
$\boldsymbol \nu_{\rm max}$ (solid) and
$\boldsymbol \nu_{\rm mean}$ (dashed),
(b) same for counts of recovered  and active cases,
(c) SIJR simulations of the dynamics of diagnosed recovered, dead, active 
and total cases for $\boldsymbol \nu_{\rm max}$ (solid) and
$\boldsymbol \nu_{\rm mean}$ (dashed).
Histograms representing (d) a discrete approximation to the probability 
distribution of parameters and (e) probabilities for  the total number of 
people affected by the virus at the end of the period.
The affected people are $25327$ for $\boldsymbol \nu_{\rm mean}$  
and  $20465$ for $\boldsymbol \nu_{\rm max}$.
Sampling parameters $W=500$, $S=4 \times 10^6$, $B=S/4$, and 
acceptance parameter $a=2$.  
}
\label{fig2}
\end{figure}

The initial stage of the epidemic corresponds to spread in the absence of any 
contention measures, see data reproduced in Fig. \ref{fig2}(a)-(b). Our goal here is 
to first fit the coefficients of the models to such data with quantified uncertainty
and then estimate a range of values for the total number of affected individuals at 
the end of the period, including exposed and undiagnosed infected individuals.

We use the guess obtained in Section \ref{sec:fitting} as a mean for the prior 
distribution (\ref{prior}), that is,
\begin{eqnarray}
\boldsymbol \nu_0= (t_{\rm in,0}, \beta_0,  \gamma_{2,0}, 
\delta_0, \alpha_0,  \ell_0), \quad 
\gamma_{1,0}^{-1} = \gamma_{2,0}^{-1}  + \alpha_{1,0}^{-1}.
\label{nu0}
\end{eqnarray}
For the different rate parameters, the deviations $\sigma_i$ will not be large.
In the absence of a better insight we can take $\sigma_i=0.1,$ 
$i=2,..,n,$ for instance. The first day of the outbreak is subject to the 
largest variance. We usually set $\sigma_1=10$.  For the likelihood (\ref{likelihood}), 
we set $M=20$ (first $20$ days) with deviations $\sigma_J = \sigma_R = 10^3$
and $\sigma_D = 10^2.$ We then sample the posterior distribution (\ref{posterior})
by MCMC techniques \cite{hammer}.
Sampling is initialized with $W$ walkers drawn from the prior distribution, 
which generate $W$ chains mixed during $K$ steps depending on an acceptance
parameter $a$. Discarding the first $B$ samples produced (to account for the 
so-called burn in period), we use the remaining
$S=KW-B$  samples to draw histograms representing the marginal probabilities
of the different model parameters, see Figure \ref{fig2} (d). 
We set $\boldsymbol \nu_{\rm max}$ to be the sample with largest posterior 
probability and $\boldsymbol \nu_{\rm mean} $ the mean of the parameter
samples, see Table \ref{table2}.

\begin{figure}[!h]
\centering
(a)   \hskip 6cm (b) \\
\includegraphics[width=6.8cm]{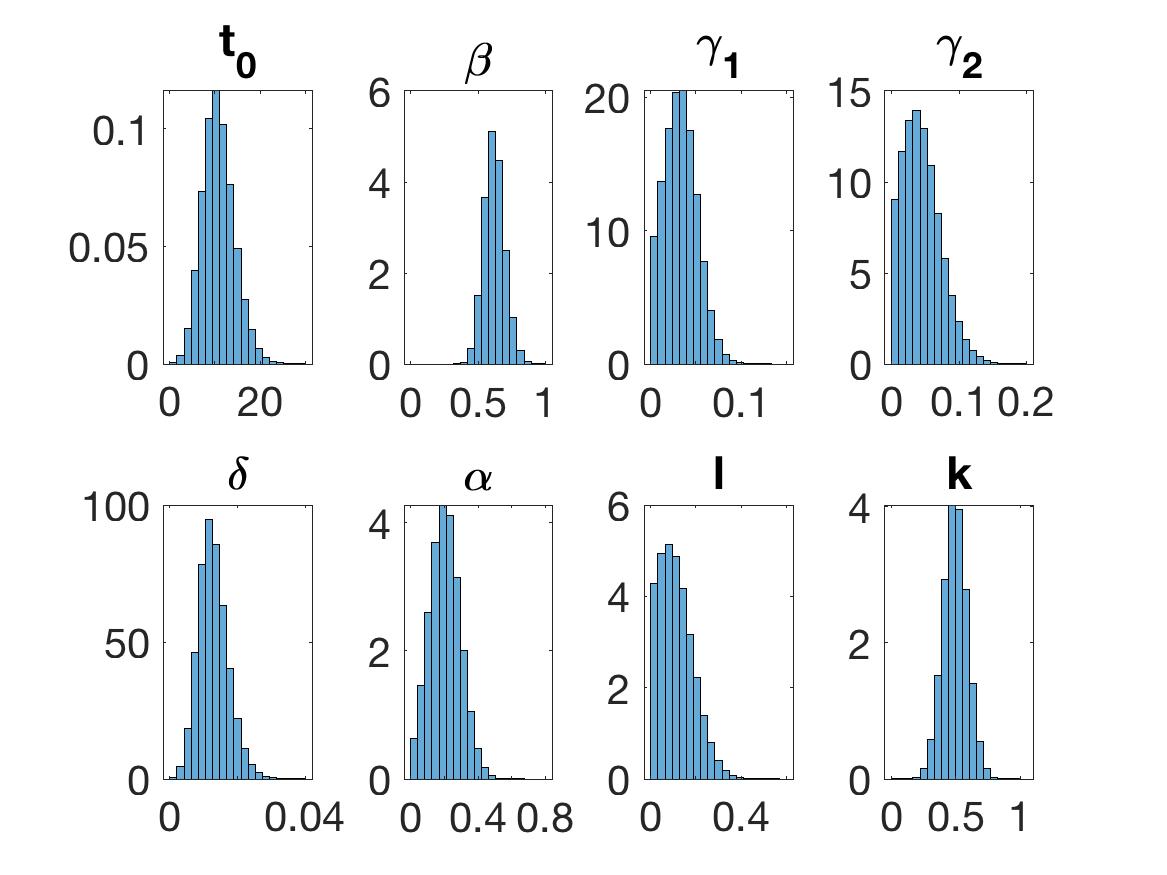}  \hskip -5mm  
\includegraphics[width=6.8cm]{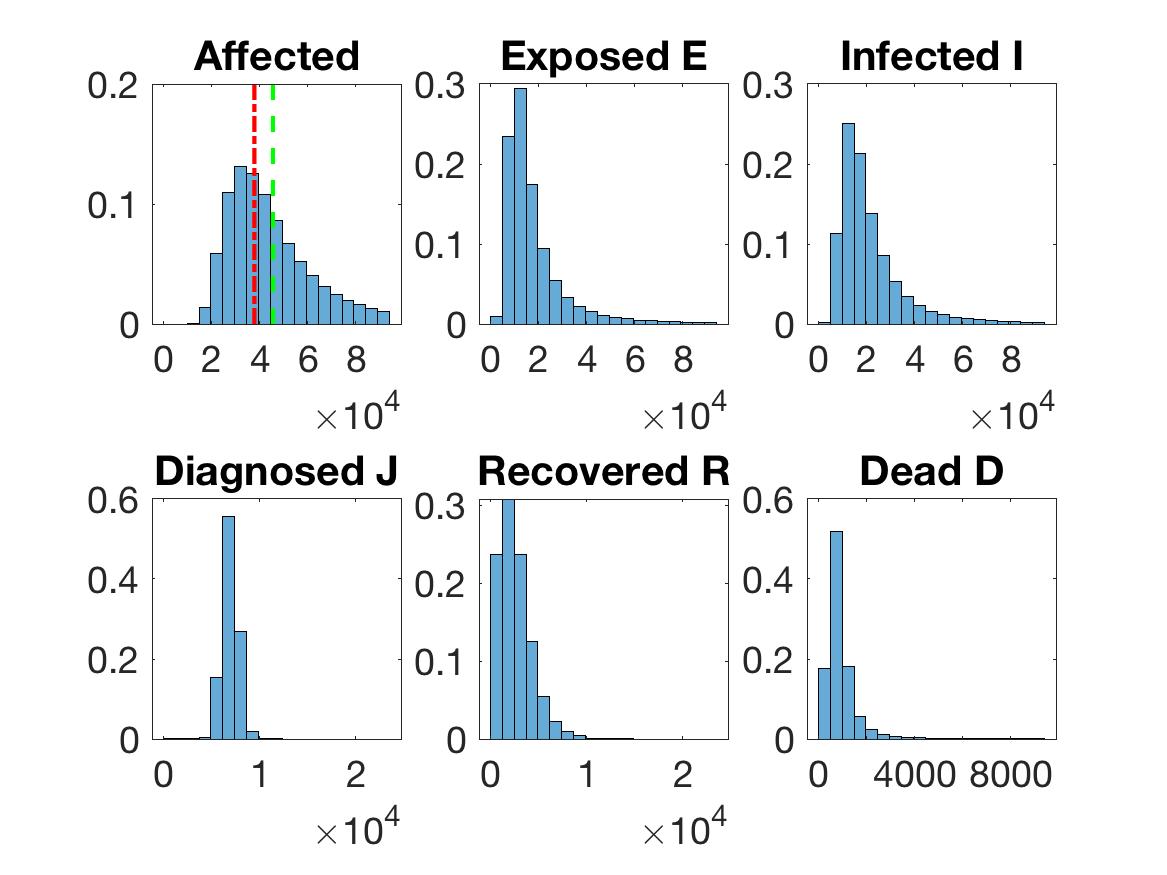}  \\
(c) \hskip 3.5cm (d) \hskip 3.5cm (e) \\ 
\includegraphics[width=4.8cm]{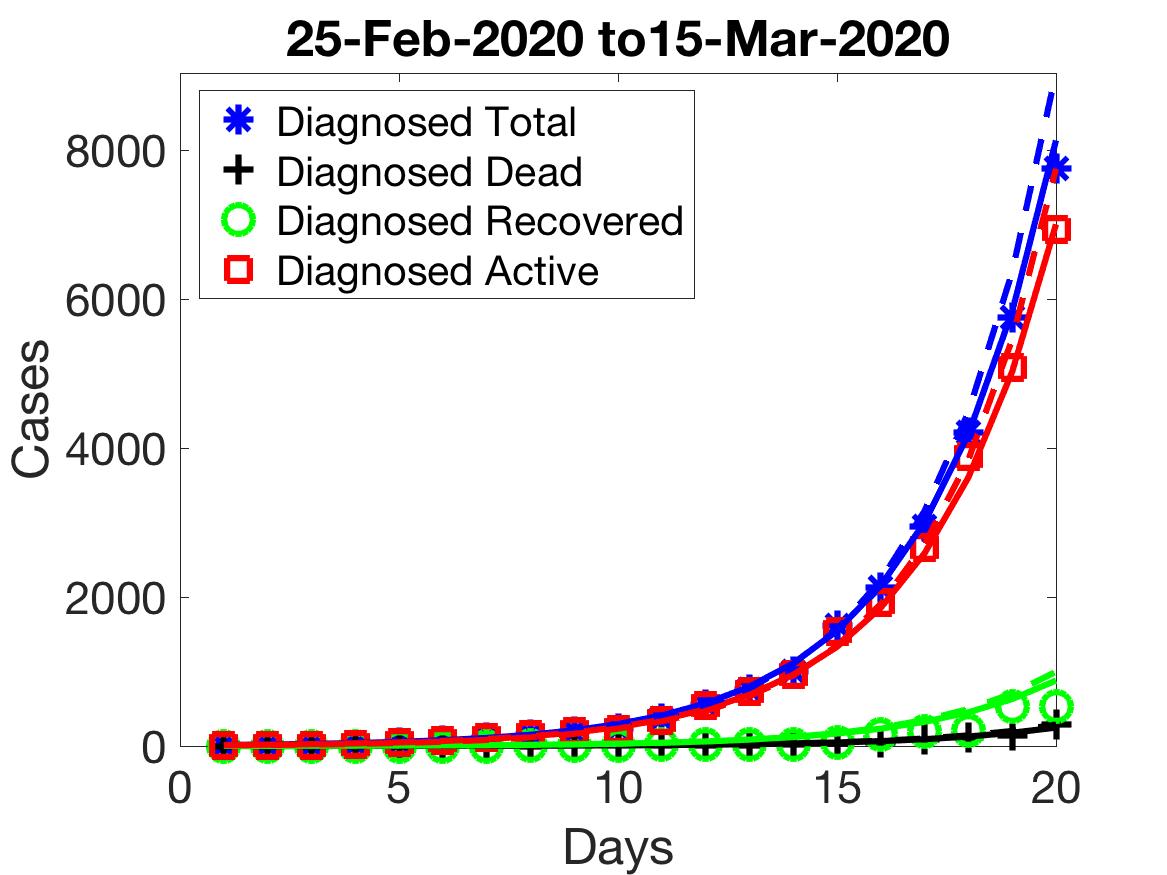}    \hskip -5mm  
\includegraphics[width=4.8cm]{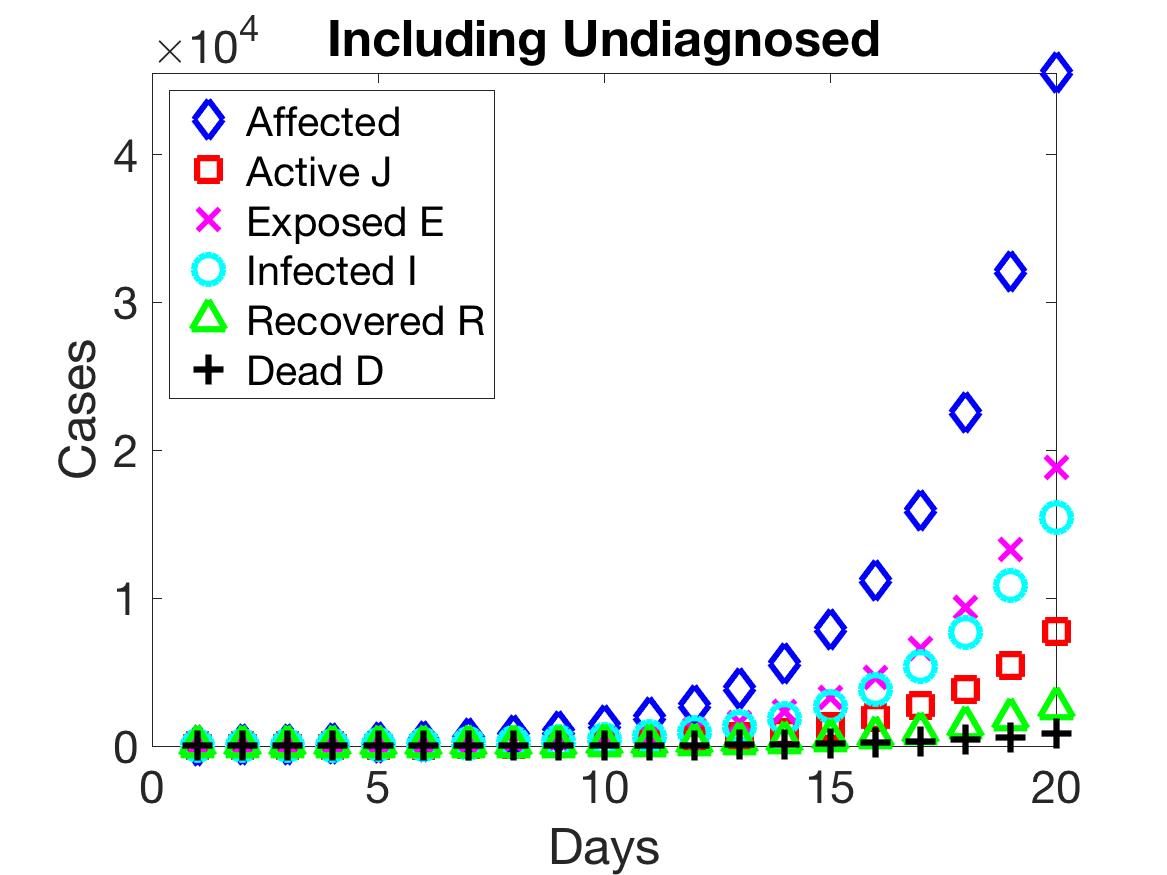}      \hskip -5mm  
\includegraphics[width=4.8cm]{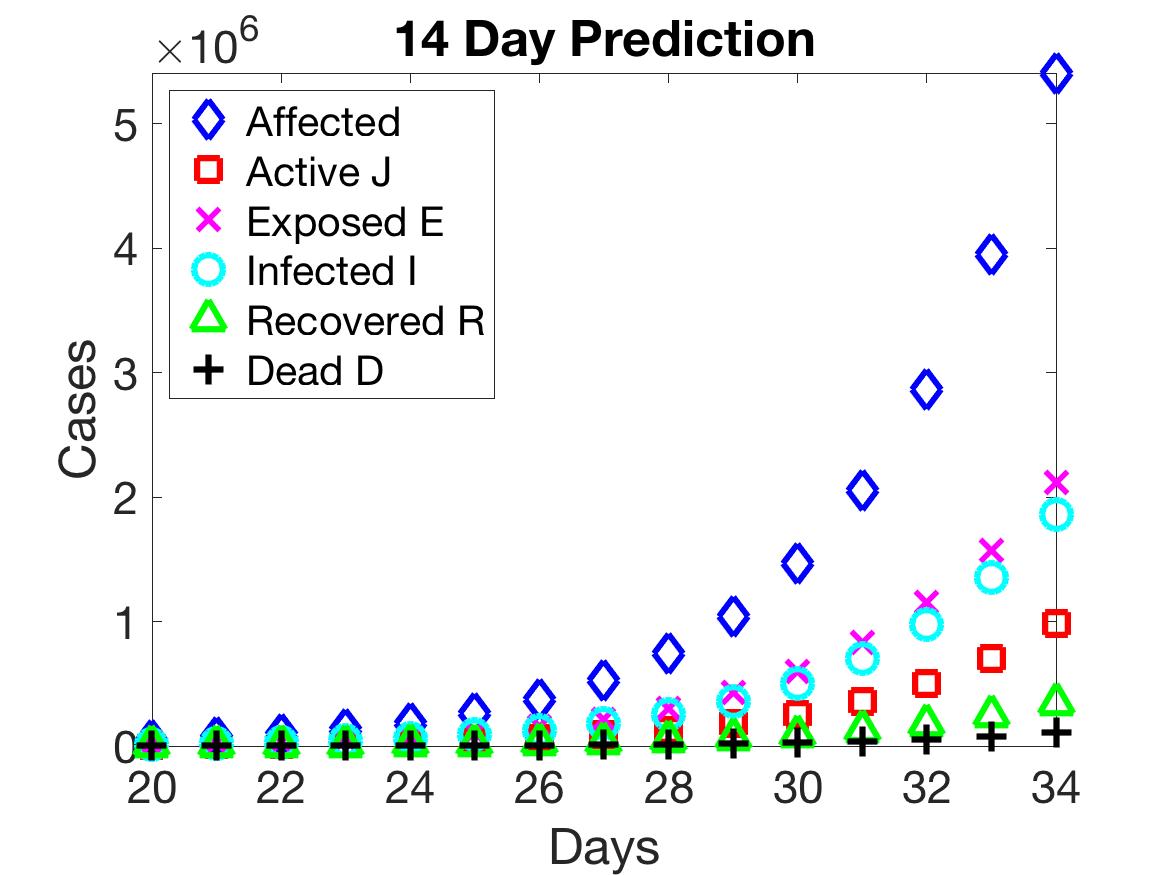}  
\caption{Initial stage using the SEIJR model:
Histograms representing (a) a discrete approximation to the probability distribution 
of some parameters and (b) the probability of different populations at the 
end of the period, including the total number of people affected by the virus at 
that time.  The affected people are $45515$  for $\boldsymbol \nu_{\rm mean}$ 
(green dashed line) and $37812$  for $\boldsymbol \nu_{\rm max}$ (red 
dot-dashed line). 
Data for diagnosed, dead, recovered and active cases are compared to
solutions of (\ref{SEIJR}) and (\ref{RjDj}) for $\boldsymbol \nu_{\rm max}$ (solid) 
and $\boldsymbol \nu_{\rm mean}$ (dashed) in (c).
SEIJR predictions of the numbers of exposed, infective, recovered and dead for 
$\boldsymbol \nu_{\rm mean}$, including undiagnosed and asymptomatic individuals,  
are shown in (d) for the initial period and in (e) for a later time.
Sampling parameters $W=500$, $S=4 \times 10^6$, $B=S/4$, and $a=2$.  
}
\label{fig3}
\end{figure}

Derived magnitudes can be visualized through histograms too, such as the final 
number of affected people in Figure \ref{fig2} (e). It has been calculated
solving equations (\ref{ecS})-(\ref{ic}) with the samples as coefficients
and computing $A = I + J + R + D$ at the final time, $20$ days.
We have superimposed the predictions for $\boldsymbol \nu_{\rm max}$ and 
$\boldsymbol \nu_{\rm mean}$.
Note that $\boldsymbol \nu_{\rm max}$ does not have a statistical meaning,  
it keeps track of a possible best fit to the data. On the other hand,
$\boldsymbol \nu_{\rm mean}$  represents some kind of average behavior.
When the distributions under study are symmetric, it will be close to 
$\boldsymbol \nu_{\rm max}$. Otherwise, it may depart from it.  In our case,
slight asymmetry is caused by discarding negative values.
In principle, we could try to improve our estimate of the parameter values
that maximize the likelihood by optimization procedures \cite{sergei}. In practice,
enforcing the positivity constraint while doing it may be problematic, and the
best samples provide reasonable approximations for our purposes.

Panels (a)-(b) in Fig. \ref{fig2} compare the observations that would be obtained with 
$\boldsymbol \nu_{\rm max}$ and $\boldsymbol \nu_{\rm mean}$ to the original 
data. If we solve the SIJR model for a longer time, for instance, $14$ days 
more, we reach about $8-10 \times 10^5$ diagnosed individuals, see panel (c), 
and about $2.25-3.57 \times 10^6$  affected people in the absence of contention 
measures.

The number of people affected by the virus with a SIJR model $I+J+R+D$ does not 
consider exposed individuals $E$. If we wish to estimate them, we 
need to use the SEIJR model. Figure \ref{fig3} summarizes some results, quite similar
to those for SIJR except for the magnifying effect of including the exposed $E$. The number  of affected individuals $E+I+J+R+D$ increases considerably, however the variation in $I+J+R+D$ is small: $22375$ for $\boldsymbol \nu_{\rm max}$ and $26687$ for  $\boldsymbol \nu_{\rm mean}$ instead of $20465$ and $25327$ for SIJR, respectively.  See Table \ref{table2} for a comparison of the parameter values for both models. Note the high transmission rate $\beta$ (about $0.6$) and the low diagnosis rate $\alpha$ (about $0.2$). Most infected individuals are not detected. If we solve the  SEIJR model for a longer time, for instance, $14$ days more, we reach about  $8-10 \times 10^5$ diagnosed individuals and $3.6-5.4 \times 10^6$ affected people  for $\boldsymbol \nu_{\rm max}$ and $\boldsymbol \nu_{\rm mean}$ in the absence of contention measures.

\begin{table}[!h]
\small \centering
\begin{tabular}{|c|c|c|c|c|c|}
  \hline
  & $\boldsymbol \nu_{\rm mean}$ SIJR  &  $\boldsymbol \nu_{\rm mean}$ SEIJR 
  & $\boldsymbol \nu_{\rm max}$ SIJR  &  $\boldsymbol \nu_{\rm max}$ SEIJR 
  & $\boldsymbol \nu_0$ 
  \\ \hline
$t_{\rm in} $             &   7.3786  & 10.7869  & 8.3266  & 11.3098  & 12.3388
  \\ \hline
$\beta$                    &   0.5938  & 0.6223  & 0.5890 & 0.6078 & 0.6262
  \\ \hline
$\gamma_1$           &    0.0390  & 0.0370 &  0.0321 & 0.0349 & 0.0667
  \\ \hline
$\gamma_2$           &   0.0473  & 0.0452 & 0.0372  & 0.0417 & 0.1000
  \\ \hline
$\delta$                   &  0.0135  & 0.0129 &  0.0115  & 0.0117 &  0.1000
  \\ \hline
$\alpha$                  &  0.2230  &  0.2051 & 0.2366 & 0.2161 & 0.2000
  \\ \hline
$\ell$                       &  0.1104  &  0.1138 & 0.0625 & 0.0694 & 0.0714
  \\ \hline
$q$                          &    &  0.4947  &   &  0.4975    &  0.5000
  \\ \hline
$k$                          &    & 0.4966 & & 0.4809   &  0.5000
  \\ \hline
$\log(p_{\rm post})$  &  -2.1343  &    &  -1.1836  &    &  -119.2493 
  \\ \hline
$\log(p_{\rm post})$  &    &  -1.9204  &     &  -1.0640  &  -58.1152 
 \\ \hline
\end{tabular}
\caption{Values of  $\boldsymbol \nu_{\rm mean}$ and $\boldsymbol \nu_{\rm max}$ 
during the initial stage using the SIJR and SEIJR models.
}
\label{table2}
\end{table}

In the next section we study the influence of contention measures on the 
subpopulations by means of the SEIJR model distinguishing two populations, 
one of which is confined.

\section{Incorporating the effect of contention measures}
\label{sec:contention}

To incorporate the effect of confinement we consider the SEIJR model
with two populations $S_1$ (unconfined) and $S_2$ (confined). During
the first period  of free growth $[t_{\rm in},T_1]$, we have $S_2=0$. The 
different periods for the data shown in Fig \ref{fig0}(a)
are marked by variations in these populations as a result of 
confinement measures. In each $i-$th period $[T_{i-1},T_i]$, $i >1$, we solve
the SEIJR model (\ref{SEIJR}) using as initial values the final values  from 
the previous period at $T_{i-1}$, for all the variables except for $S_1$ 
and $S_2$:
\begin{itemize}
\item Period 2: $(1-\rho) S_1(T_1) $ and $S_2(T_1) +
\rho S_1(T_1) $ are used as initial data for $S_1$ and $S_2$, 
respectively. 
\item Period 3: $(1-\rho) S_1(T_2) $ and $S_2(T_2) +
\rho S_1(T_2) $ are used as initial data for $S_1$ and $S_2$, 
respectively.
\item Period 4: $S_1(T_3) + (1- \rho) S_2(T_3)$ and 
$\rho S_2(T_3)$ are used as initial data for $S_1$ and $S_2$, 
respectively.
\end{itemize}
Recall that in the first period, the initial values for all the
variables are zero, except $E(t_{\rm in})=1$ and $S_1(t_{\rm in})=N-1.$
No parameter $t_{\rm in}$ appears in the next periods, we set it equal
to zero.  Instead, we introduce $\rho \in (0,1)$ to quantify  the abrupt changes 
in the fraction of people confined at the start of each period.
We assume that the transmission rate for $S_2$ is lower by a factor $p$,
that is, $p \beta$ instead of $\beta$, due to the reduction of contacts with 
other people. Due to possible interaction with already sick people or people 
still working outside at home, we cannot set it equal to zero.

\begin{figure}[h!]
\centering
(a) \hskip 3.5cm (b) \hskip 3.5cm (c)  \\
\includegraphics[width=4.8cm]{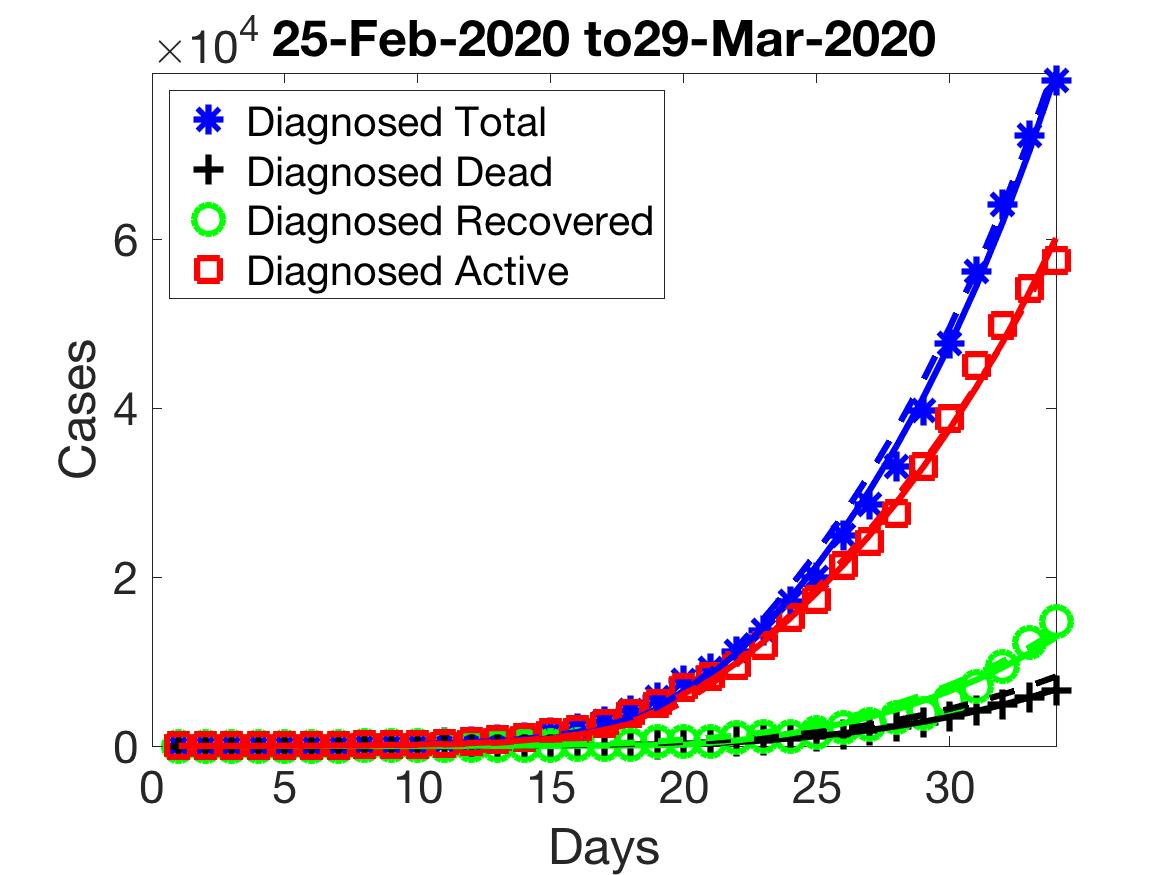} \hskip -5mm
\includegraphics[width=4.8cm]{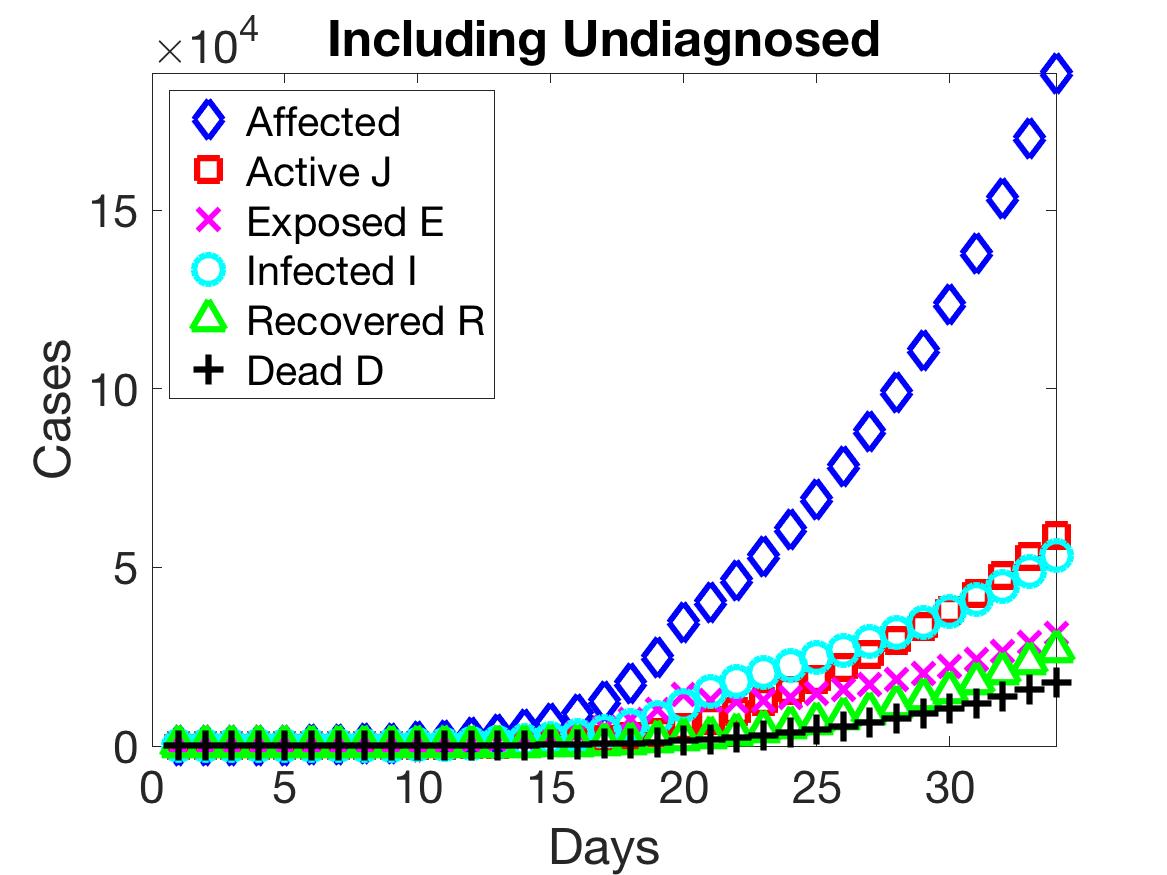} \hskip -5mm
\includegraphics[width=4.8cm]{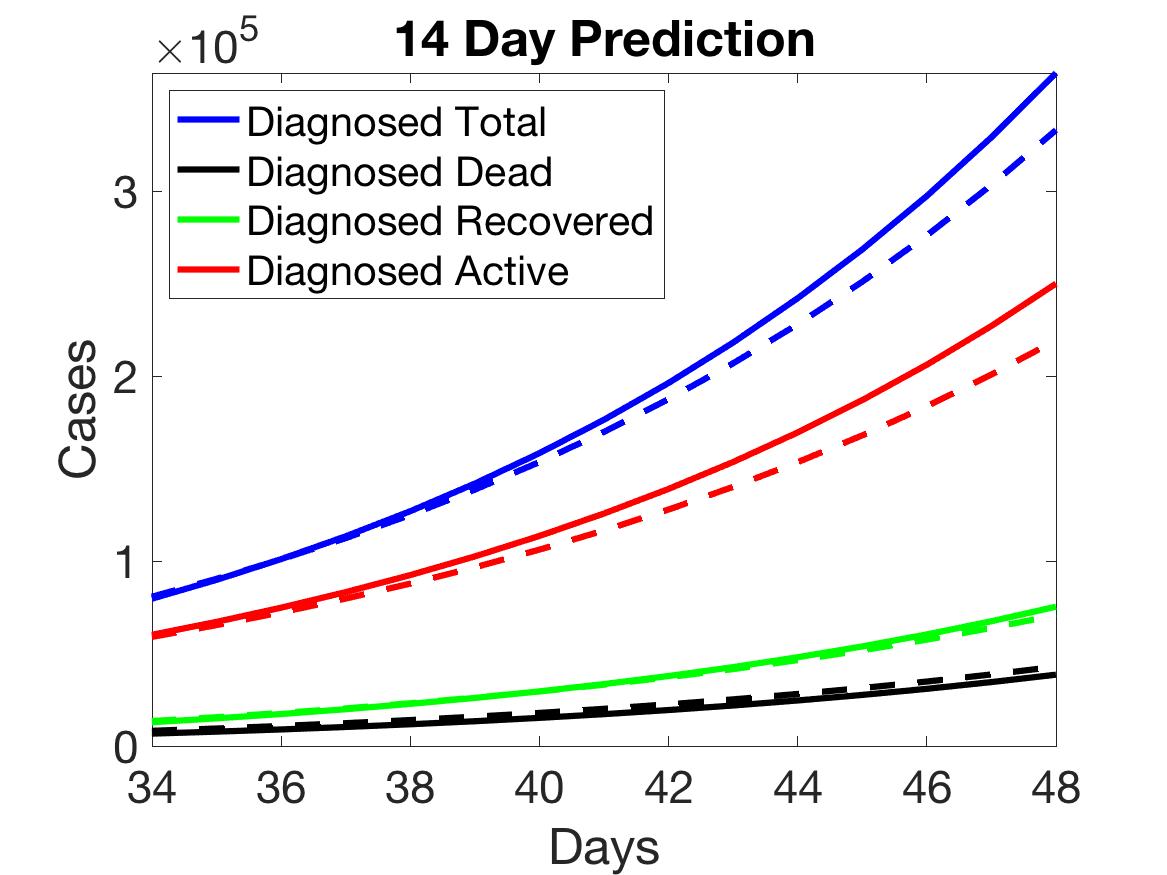}  \\
(d) \hskip 6cm (e)  \\
\includegraphics[width=6.8cm]{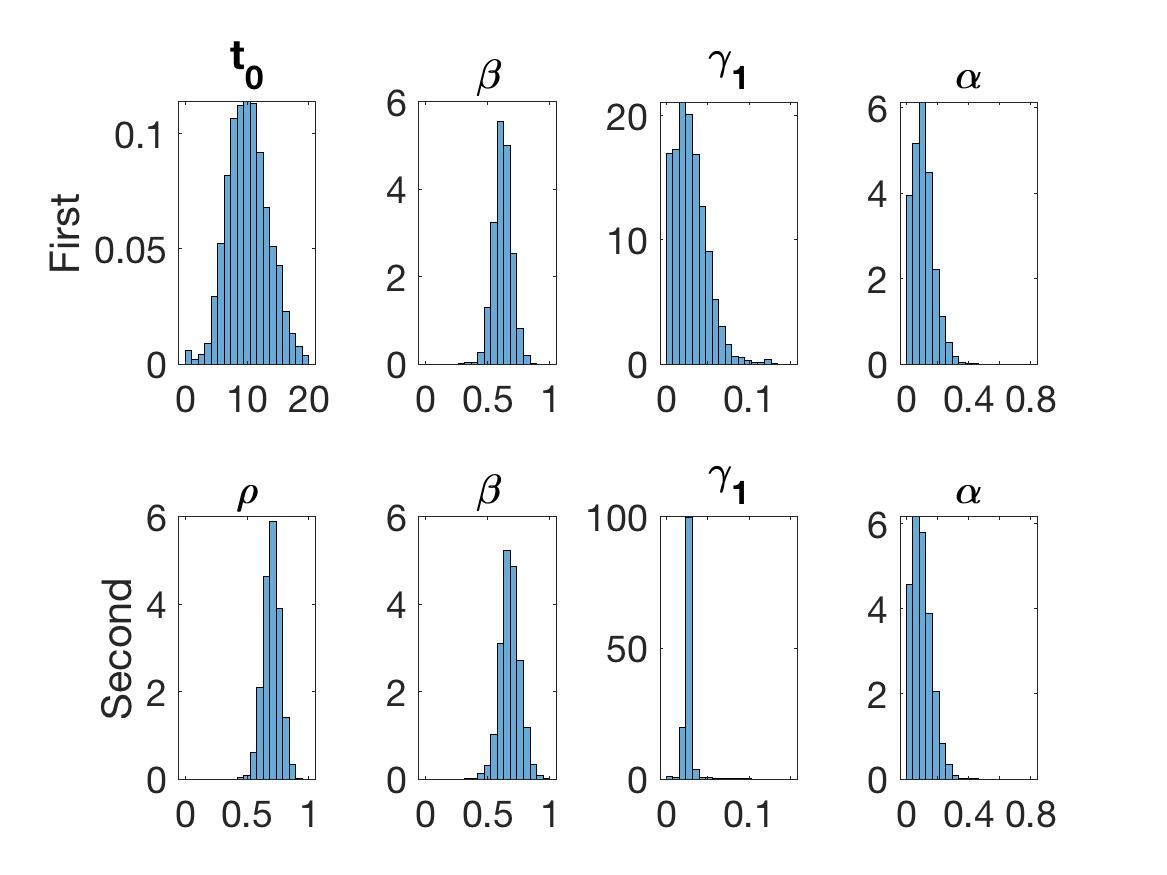}  \hskip -5mm 
\includegraphics[width=6.8cm]{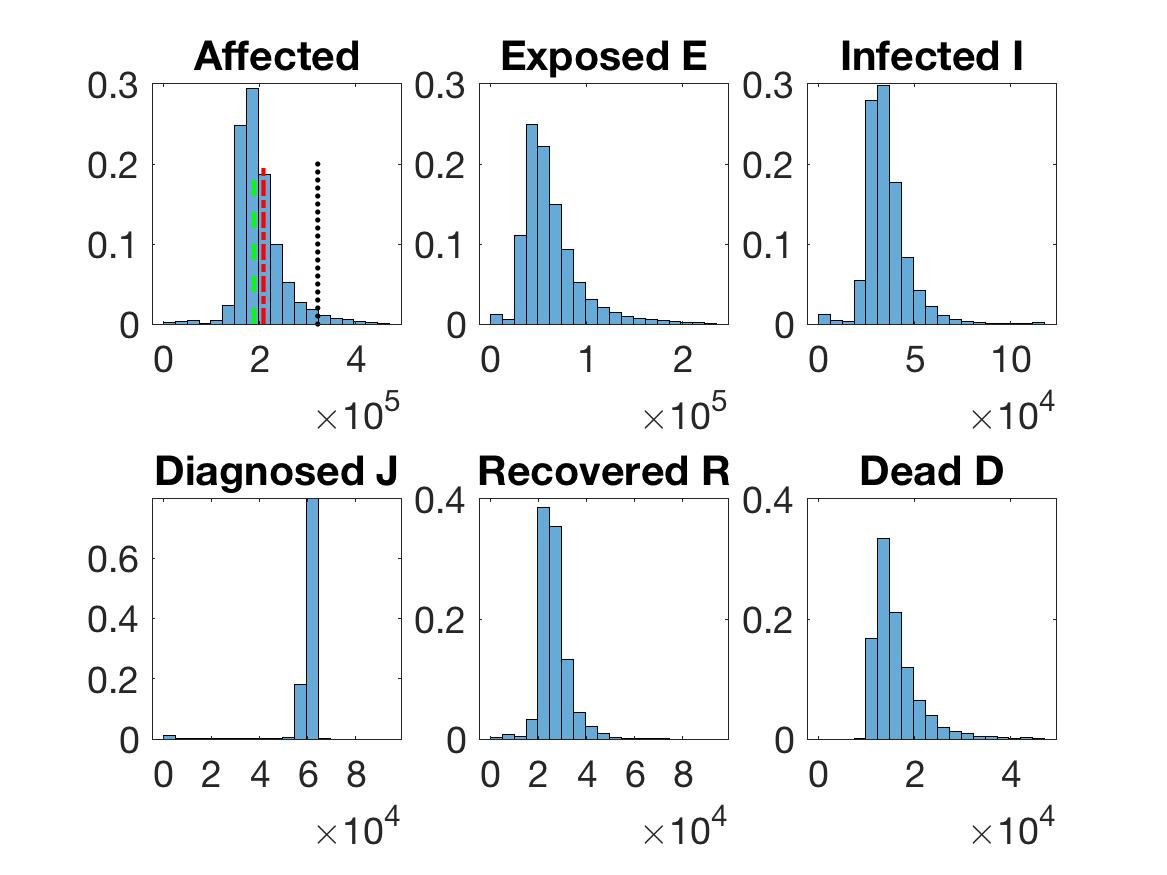}  
\caption{First and second periods:
(a) Data for diagnosed (asterisks), dead (crosses), recovered (triangles) and 
active (squares) cases,  compared to solutions of (\ref{SEIJR}) and (\ref{RjDj}) 
for $\boldsymbol \nu_{\rm max}$ (solid) and $\boldsymbol \nu_{\rm mean}$ 
(dashed), extended in (c) for a longer time.
(b) SEIJR simulations of the numbers of  exposed, infective, recovered and dead individuals for $\boldsymbol \nu_{\rm mean}$ including the undiagnosed and
asymptomatic.
(d) Histograms comparing the distribution of some parameters in the two periods.
(e) Histograms representing the probability of the status of different populations 
at the end.   Vertical lines mark the values for 
$\boldsymbol \nu_{\rm max}$ (red dot-dashed line), 
$\boldsymbol \nu_{\rm mean}$ (green dashed line),
and the mean for all samples (black dotted line).
Sampling parameters $W=500$, $S=4 \times 10^6$, $B=S/4$ and $a=2$. 
}
\label{fig4}
\end{figure}

We adapt the framework presented in Section \ref{sec:bayesian} ensambling 
these periods as we explain next. To consider stages $i$, $i=1,\ldots,Q$, we 
multiply the number of parameters by $Q$. The first block of $9$ parameters is the 
standard one for the first period. The remaining blocks correspond each to one 
additional period, with $t_{\rm in}$ replaced by $\rho$. We keep the same initial 
guesses of the parameters used in  Section \ref{sec:free} as prior knowledge 
in all the  periods, except for $\rho$, which is set equal to $3/4$, $4/5$, $15/16$ respectively, an approximation of the population switches at the different stages. 
The deviations are kept equal to $0.1$ for all, except $t_{\rm in},$ for which we 
set it equal to $10$. As for the data, we keep the same deviations as in
Section \ref{sec:free} in all the periods, in the absence  of better information.

Let us consider first the initial confinement period.
Figure \ref{fig4}(a) compares to data the evolution of the diagnosed subpopulations.
Population dynamics is calculated solving the SEIJR model in two sequential steps, 
in $[t_{\rm in},T_1]$ and $[T_1,T_2]$, using in each of them the parameter 
values obtained for that period and the initial data stipulated earlier.  Panel (b)
represents the solutions of the SEIRJ model including the contribution of undiagnosed
and asymptomatic individuals. Panel (d) compares the distribution of
some parameters in the two periods. The transmission rate $\beta$ increases slighty
in the second period, while the diagnose rate remains low. These histograms are discretizations of the probability, so that the height of each bin is the number of samples divided by the total and by  the basis of the bins (which is the same for the histograms corresponding to the same parameters in this figure and the previous ones to allow for comparisons). Figure \ref{fig4}(e) quantifies uncertainty in the total number of people affected by the virus after these two periods. 
If we keep the parameter values  $\boldsymbol \nu _{\rm max}$ or $\boldsymbol \nu _{\rm mean}$ up to time $T_3>T_2$, growth slows down, but it does not stabilize, see Fig. \ref{fig4}(c). 

\begin{figure}[h!]
\centering
(a) \hskip 5.5cm (b)  \\
\includegraphics[width=6cm,valign=c]{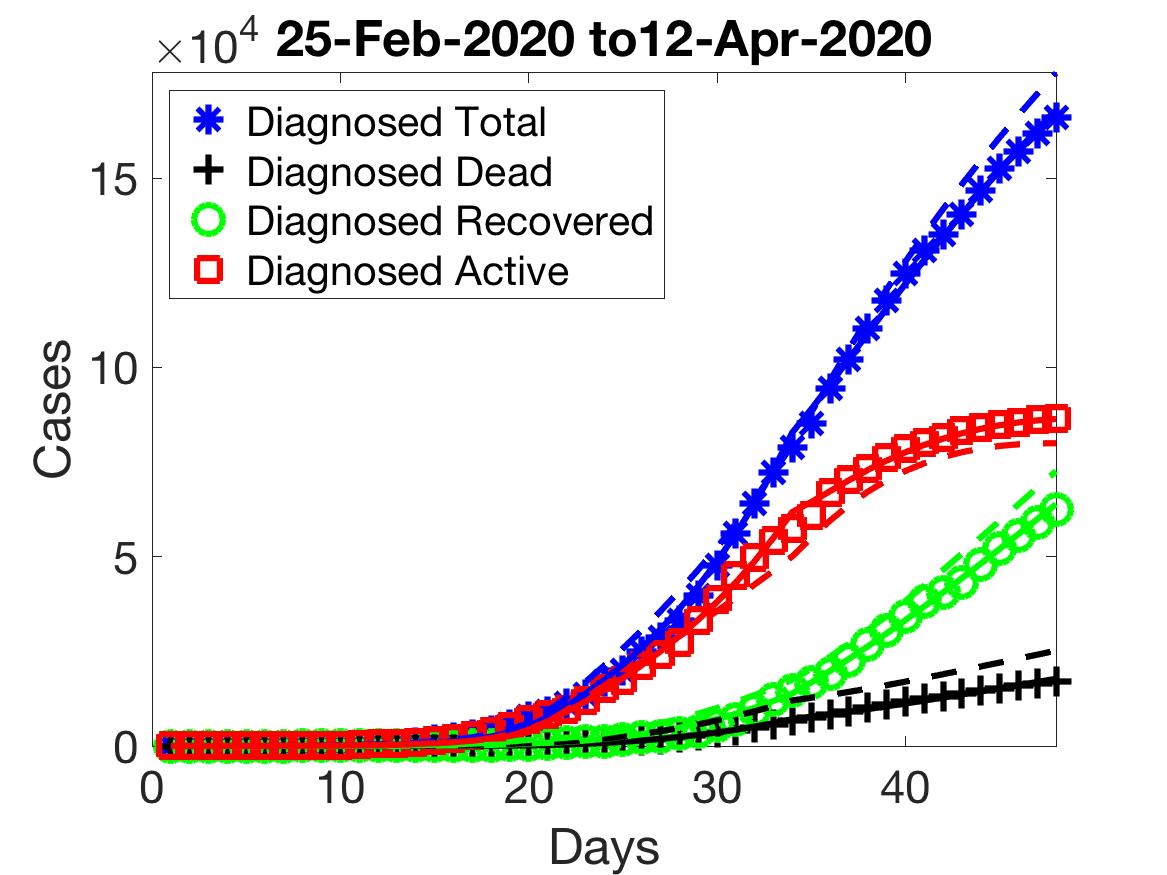}
\includegraphics[width=6.8cm,valign=c]{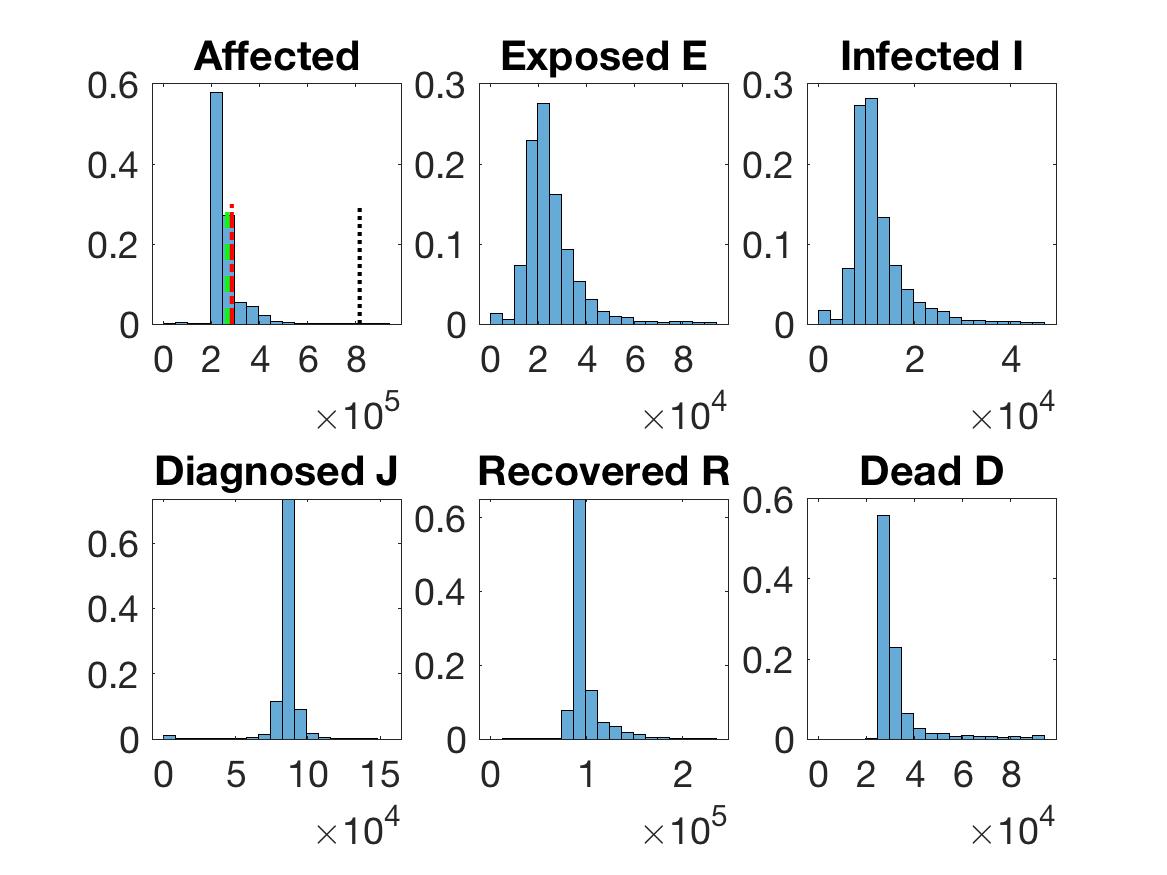}\\
\caption{Same as Fig. \ref{fig4}(a) and (e) for three periods of increasing
confinement. Note the decrease in the numbers of exposed $E$ and infected 
$I$ cases.
Sampling parameters $W=500$, $S=5 \times 10^6$, $B=S/4$ and $a=2$. 
}
\label{fig5}
\end{figure}

We incorporate next the third additional period in which an even larger fraction of the
population is confined at home. The results are reproduced in Figure \ref{fig5}. Finally,
the growth trend moderates, also in predictions for longer times, see Fig \ref{fig1}(b).
Table \ref{table5} reports the mean values $\boldsymbol \nu_{\rm mean}$ obtained 
after MCMC sampling, as well as the values corresponding to the best sample 
$\boldsymbol \nu_{\rm max}$. As mentioned earlier, $\boldsymbol \nu_{\rm max}$
has not a statistical meaning. It represents a best fit whose coefficients may fluctuate
a bit with the number of samples. Instead, $\boldsymbol \nu_{\rm mean}$ conveys
a statistical trend of the coefficients of the samples. Comparing the values of $\beta$ 
for $\boldsymbol \nu_{\rm mean}$, we remark an increase in $\beta$ in the second
period. This fact is also observed in $\boldsymbol \nu_{\rm max}$ and the trend was already present in the histograms for $\beta$ in Fig \ref{fig4}(d).
According to the information available on the spanish outbreak, and taking into account that infected people can take up to $14$ days to show symptoms, this might be a delayed reflection of crowd gatherings occurred at the end of the first period, or also, a result of the lack of protective equipment for overwhelmed health care and security workers.
We also observe a reduction in the mean recovery rates for $\gamma_1$ and 
$\gamma_2$ in the second period, which may be reflection of the saturation of the 
health care system and the  scarceness of medical  resources during the second period. 
Notice that the diagnose rate $\alpha$ is quite low. A large fraction of affected
people remains undetected.

\begin{table}[!h]
\small \centering
\begin{tabular}{|c|c|c|c|c|c|c|}
  \hline
  & $\boldsymbol \nu_{\rm mean}$ 1st & $\boldsymbol \nu_{\rm mean}$ 2nd &   
  $\boldsymbol \nu_{\rm mean}$ 3rd
  & $\boldsymbol \nu_{\rm max}$ 1st & $\boldsymbol \nu_{\rm mean}$ 2nd &  
  $\boldsymbol \nu_{\rm max}$ 3rd \\
  \hline
$t_{\rm in}$, $\rho$   &  12.3479 &  0.7202 &  0.2236  &  7.9770  &  0.8064 &  0.2527
\\  \hline
$\beta$  &   0.6173 &  0.6902 &  0.5898 &  0.6894 &  0.7028  &  0.6796
\\  \hline
$\gamma_1$ &   0.0741 &  0.0363  &  0.0446 &  0.0410 &  0.0290 &  0.0318
\\  \hline
$\gamma_2$ &  0.1426  &  0.0437  &  0.0551 &   0.0532  &  0.0343  &  0.0461
\\  \hline
$\delta$ &   0.0696  &  0.0310  &  0.0131  &  0.0141  &  0.0180  &  0.0098
\\  \hline
$\alpha$ &   0.1541  &  0.2148  &  0.2343  &  0.1791  &  0.1851  &  0.1022
\\  \hline
$\ell$ &   0.1056  &  0.1245  &  0.1031  &  0.1253  &  0.1800  &  0.0219
\\  \hline
$q$ &  0.4978  &  0.5301  &  0.5082  &  0.3872  &  0.7778  &  0.4234
\\  \hline
$p$ &     &  0.1139  &  0.0643  &   &  0.1845  &  0.0001
\\  \hline
$k$ &  0.4984  &  0.5106  &  0.5251  &  0.5802  &  0.5703  &  0.5339
\\ \hline
\end{tabular}
\caption{Values of   $\boldsymbol \nu_{\rm mean}$ and $\boldsymbol \nu_{\rm max}$ 
for three periods using the SEIJR model, with 
$\log(\boldsymbol \nu_{\rm mean}) =  3.2974$, 
$\log(\boldsymbol \nu_{\rm max}) =  188$, 
$\log(\boldsymbol \nu_{0})= -71097$, respectively. In the first row, the first columns
represent $t_0$, while the rest correspond to $\rho$.
}
\label{table5}
\end{table}

\begin{figure}[h!]
\centering
(a) \hskip 5.5cm (b)  \\
\includegraphics[width=6cm,valign=c]{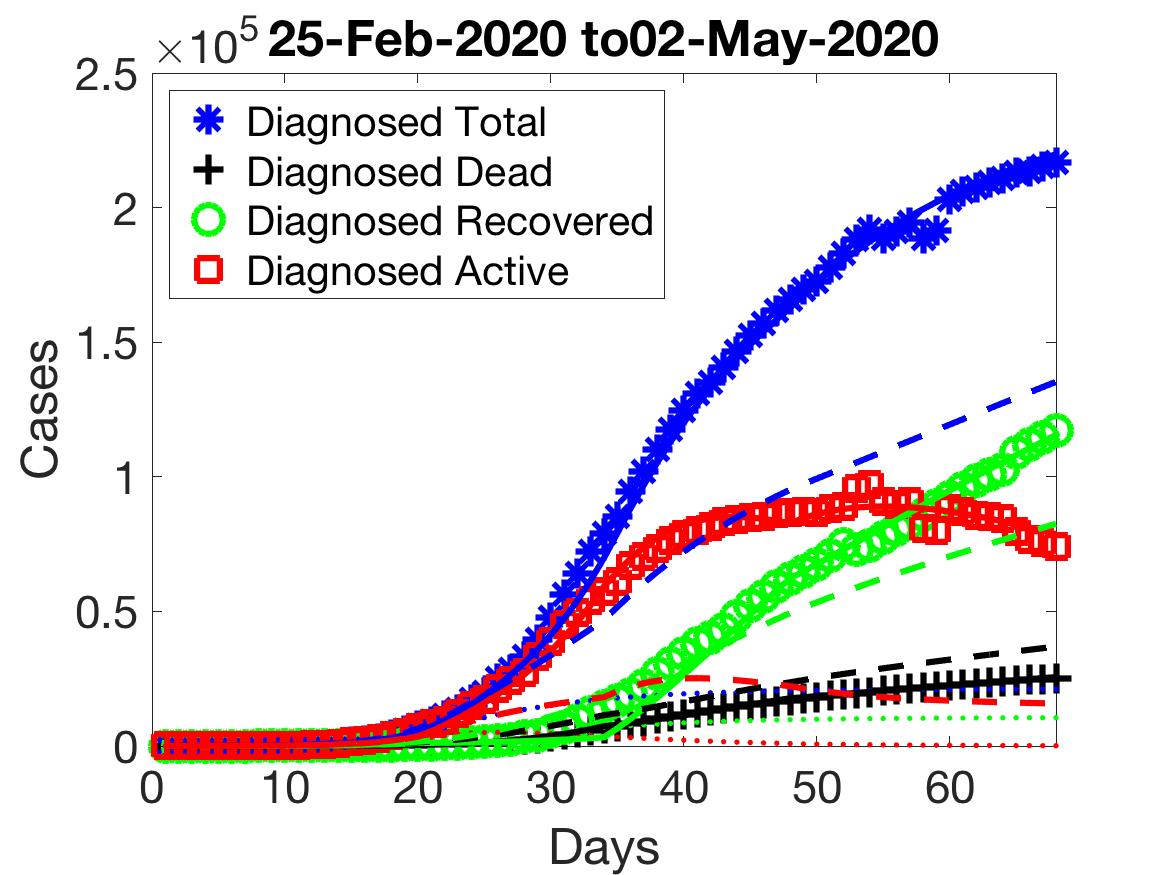}
\includegraphics[width=6.8cm,valign=c]{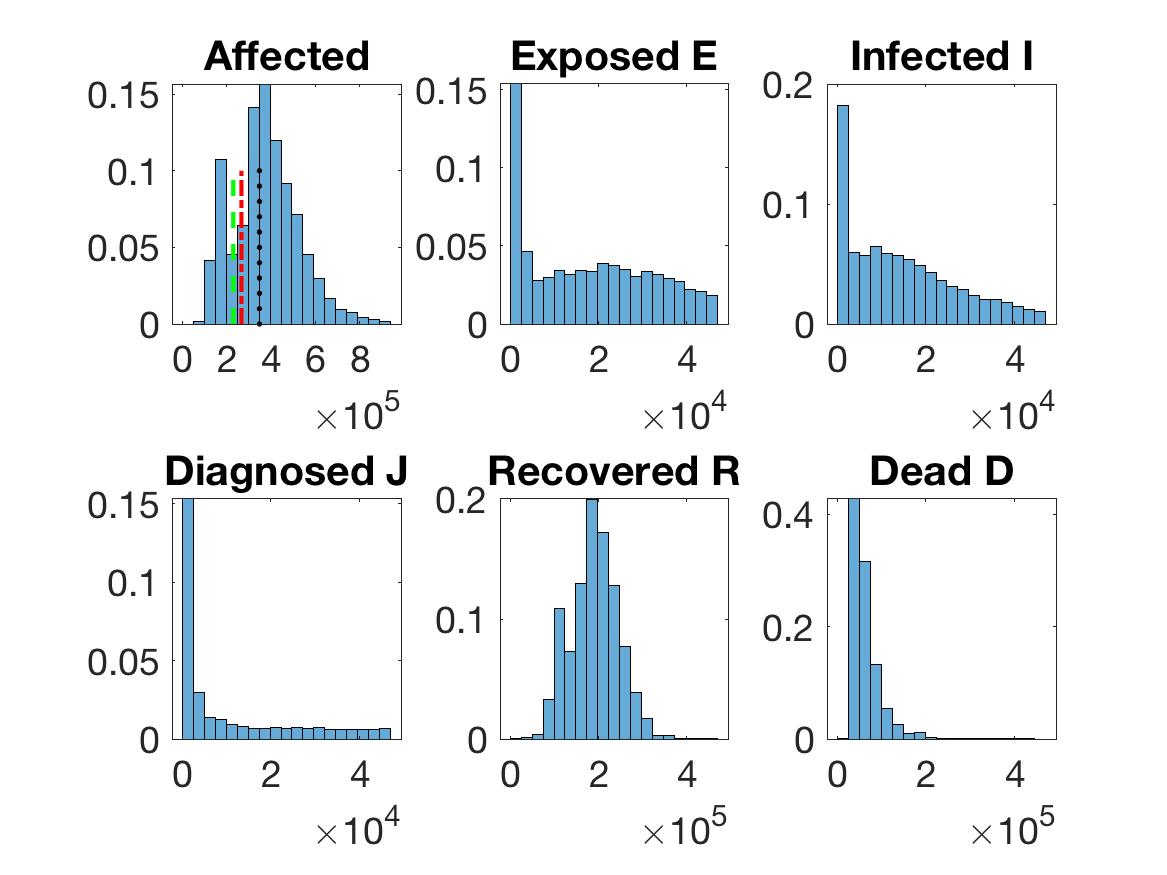}\\
\caption{Same as Fig. \ref{fig5}(a)-(b) for four periods. Additional dotted lines in the
lower part of panel (a) represent solutions of (\ref{SEIJR}) and (\ref{RjDj}) for 
$\boldsymbol \nu_0$. The numbers of exposed $E$, infected $I$ and active diagnosed
$J$ individuals are depleted.}
\label{fig6}
\end{figure}

In a fourth period, a fraction of the population is released from confinement.
The number of undiagnosed and exposed individuals is depleted and
the spread of the epidemic is contained.
Unlike before, the SEIJR solutions for $\boldsymbol \nu_{\rm max}$ still fit
the data quite well, but the solutions for $\boldsymbol \nu_{\rm mean}$
deviate from the data towards the solution for the prior $\boldsymbol \nu_0$,
see Fig. \ref{fig6}(a). This reflects some kind of bimodality, with a collection
of SEIJR solutions  close to the prior while most of them remain close to the 
data as the model coefficients  range through the sampled parameters.
This may be a consequence of fixing prior guesses for the model parameters 
that worsen  with time. Note that the predictions that would be obtained 
using the prior are rather poor,  compared to true counts, as time grows.
However, the predictions provided by  $\boldsymbol \nu_{\rm max}$ fit
the data quite well, even for later times, see Fig. \ref{fig1}(b).

Note that as we add data from new periods, we are including more information
in the analysis. The best coefficient values estimated for  previous periods
change slightly and  we infer more moderate numbers of affected individuals, 
as compared with the previous studies done  using less data. However,
the same trends persist: increase of $\beta$ in the second period, while
$\gamma_1$ and $\gamma_2$ decrease, decrease of $\delta$ and low
diagnose rate $\alpha$. Very few tests were done during these periods. In fact,
the usefulness of tests would be to increase the diagnose rate, augmenting
the number of  quarantined infected and asymptomatic individuals.

Figure \ref{fig1}(b) provides a global view of our analysis. Shaded areas
represent the total number of diagnosed cases $J+D_J+R_J$ obtained
solving (\ref{SEIJR}) and (\ref{RjDj}) for the last 1000 sampled parameters 
in each of the four frameworks we have considered: red for Period 1, green
for Periods 1-2, blue for Periods 1-2-3, magenta for Periods 1-2-3-4.
Dotted lines represent the mean of the curves obtained for all the samples.
Thicker lines represent the total number of diagnosed cases $J+D_J+R_J$ 
for $\boldsymbol \nu_{\rm max}$ (solid), $\boldsymbol \nu_{\rm mean}$ (dashed)
and $\boldsymbol \nu_0$ (dash-dotted). Yellow circles represent the data: total
counts of diagnosed people (dead, recovered and active). Colored triangles
separate the 'inference' from the 'prediction' regions for each of them.
At the back of the triangles, we have the inference region, corresponding to
the data we use to infer the parameter values and the total number of affected 
people. At the front  of the triangles, we use model solutions
to predict the time evolution keeping the conditions of the last period
considered in the inference studies. Taking no measures leads to the evolution
represented in red. Confining people who are able to work online or do not
work in basic activities results in the dynamics marked in green. Extending
the confinement to all the population not working in strictly essential activities
leads to the forecast painted in blue. Releasing this last fraction of the
population results in the evolution represented in magenta. Notice that
the solid magenta curve corresponding to $\boldsymbol \nu_{\rm max}$
agrees very well with the data past day $68$ (last day used to calculate it),
whereas some magenta samples deviate considerably. This fact is reflected 
in the dotted averages, which define somehow a confidence region. 
After day $68$ the population was released from confinement by stages,
and the use of masks was enforced, lowering the risk for the users.
The country remained closed.
The different predictions associated to the four inference studies we carried out 
are not only due to the confinement or the release of population fractions, but 
to the fact that we allow for variations in the model coefficients in the different 
periods to adapt them to additional amounts of data. The fact that the transmission coefficient $\beta$ decreases with time due to improved conditions is fundamental. 



These studies are limited by the data quality. As mentioned earlier, the order of 
magnitude of the population counts in official records changes noticeably when only 
PCR confirmed cases are taken into account or also probable cases are included. 
In the spanish outbreak, the number of probable cases may have been five times higher
and the number of dead individuals twice as much. Repeating our previous studies 
scaling the data in that way, we find estimates about 2 million people, 
consistent with the official conclusions inferred from selected testing campaings.

\section{Conclusions}
\label{sec:conclusions}


The attempt to devise mathematical models to study the progression
of a pandemic faces the need to handle large uncertainty in the available 
data. We have developed a Bayesian framework to quantify uncertainty
in the effects of lockdown measures through the coefficients of SEIJR and 
SIJR models for human-to-human transmission. A key idea is the
introduction of two populations, one of which has a lower risk of infection 
than the other.  Lower risk may be due to confinement, as it happens
for the data we consider here, or to preventive measures, such as the
use of masks. Therefore, our methodology is not constrained to lockdown
measures.

These techniques allow us to calibrate important  magnitudes to forecast 
the evolution of the epidemic, such as the variation in the total number of 
affected people (including asymptomatic individuals), and could be adapted 
to  infer coefficients from data from any country. We show how enforcing
measures that deplete the number of undiagnosed and asymptomatic 
individuals, while reducing the transmission rate,  we can stop the spread.
We have focused on the data available for Spain, 
which shows well differentiated data periods according to the 
measures taken. We see that the model coefficients in each period vary 
with the circumstances. For instance, transmission rates 
may augment as a result of increased interaction and lack of protective 
measures and recovery rates may decrease as a result of scarceness of 
resources. The diagnose rate is low, resulting in large number of 
undiagnosed individuals. Performing more PCR tests increases 
the diagnose rate, allowing to  quarantine more infected and 
asymptomatic individuals.

An additional difficulty when applying this inference framework
for large periods of time (months) is the fact that uncertainty in the
observed data accumulates over time when using cumulative data. 
This poses the problem of selecting adequate variances for the analysis. 
In the absence of reliable information in that respect, we have kept them 
fixed. Calculations with daily data do not show significative differences
in the observed trends in our case.
Moreover, we have used official data for PCR confirmed patients only. 
The effect of adding probable cases, which may have been five times 
higher, would require further consideration.


SIR type models assume that recovered individuals have immunity.
This may not be the case here, thus additional studies taking this
factor into account would be advisable \cite{susceptible1,susceptible2}. 
Furthermore, standard SIR type models \cite{covid2} are formulated
for closed systems. Introducing spatial mobility \cite{spatial,covid2} is 
an important issue that should be a subject for future work.
Moreover, imperfect implementation of contention measures leads to 
delays, which might be better described by differential-delay models 
\cite{sars_delay}.
We have focused on human-to-human transmission here. Coronaviruses 
originate in animals, such as bats, and arrive to humans through 
intermediate animal species which act as reservoirs for future waves 
\cite{covid_bat}, subject deserving further studies.

\section{Appendix: Solutions of the SIJR model}
\label{sec:explicit}

Let us obtain explicit expressions for the solution of the (\ref{ecS})-(\ref{ic})
model. Consider the equations (\ref{ecI})-(\ref{ecJ}) for $I$ and $J$.
Set $D_1 = \alpha + \gamma_1 + \delta$, $D_2 = \gamma_2 + \delta.$
The system matrix is 
\begin{eqnarray*}
A = \left( \begin{array}{cc}
\beta - D_1 & \ell \beta, \\
\alpha & - D_2,
\end{array} \right)
\end{eqnarray*}
with eigenvalues
\begin{eqnarray*}
\lambda_1 = {\beta - D_1 - D_2 \over 2} - {1 \over 2 }\sqrt{
\beta^2 - 2 \beta D_1 + 2 \beta D_2 + 4 \alpha \ell \beta + D_1^2 - 2 D_1D_2 + D_2^2}, \\
\lambda_2 = {\beta - D_1 - D_2 \over 2} + {1 \over 2 }\sqrt{
\beta^2 - 2 \beta D_1 + 2 \beta D_2 + 4 \alpha \ell \beta + D_1^2 - 2 D_1 D_2 + D_2^2},
\end{eqnarray*}
and eigenvectors:
\begin{eqnarray*}
\mathbf v_1 = (-\ell \beta, \, \beta - D_1 - \lambda_1),\quad 
\mathbf v_2 = (-\ell \beta, \, \beta - D_1 - \lambda_2 ).
\end{eqnarray*}
The general solution is 
\begin{eqnarray*}
( I(t), J(t) ) = z_1 \mathbf v_1 e^{\lambda_1 t}
+ z_2 \mathbf v_2 e^{\lambda_2 t}, \quad z_1,z_1 \in \mathbb R.
\end{eqnarray*}
We obtain the solutions for the initial value problem combining the solutions
with initial data $(1,0)$ and $(0,1)$. The coefficients $z_1,z_2$ for $(0,1)$
are
\begin{eqnarray*}
z_1 = -1/(\lambda_1 - \lambda_2), \quad
z_2 = 1/(\lambda_1 - \lambda_2).
\end{eqnarray*}
For $(1,0)$
\begin{eqnarray*}
z_1= (D_1 - \beta + \lambda_2)/(\beta \ell (\lambda_1 -\lambda_2)), \quad
z_2= -(D_1 - \beta + \lambda_1)/(\beta \ell (\lambda_1 - \lambda_2)),
\end{eqnarray*}
provide the solution to our problem. 

Set
\begin{eqnarray*}
c_1 = {\beta -D_1 - \lambda_2 \over \lambda_1 - \lambda_2},
c_2 = {\beta -D_1 - \lambda_1 \over \lambda_2 - \lambda_1}.
\end{eqnarray*}
Then the number of infected people is
\begin{eqnarray}
I(t) = c_1 e^{\lambda_1 t} + c_2 e^{\lambda_2 t}, \label{I}
\end{eqnarray}
and the cumulative number of infected people $I_c$ such that $I_c'=  I,$ 
$I_c(0)=0$,  is
\begin{eqnarray}
I_c(\beta,t) =    {c_1 \over \lambda_1} e^{\lambda_1 t} + 
{c_2 \over \lambda_2} e^{\lambda_2 t} - \left( {c_1 \over \lambda_1} 
+ {c_2 \over \lambda_2} \right). \label{Jc}
\end{eqnarray}
The number of diagnosed people is
\begin{eqnarray}
J(t) = {\alpha \over \lambda_1-\lambda_2} e^{\lambda_1 t} + 
{\alpha \over \lambda_2 - \lambda_1} e^{\lambda_2 t}. \label{J}
\end{eqnarray}

The cumulative number of diagnosed people $J_c$
is then the integral of this magnitude, $J_c'=J$ starting from zero
$J(0)=0$:
\begin{eqnarray}
J_c(t)= {\alpha \over \lambda_1-\lambda_2} \left({e^{\lambda_1 t} -1
\over \lambda_1} - { e^{\lambda_2 t}  -1 \over \lambda_2}  
\right). \label{Jc}
\end{eqnarray}

We can now integrate the equations for $S$, $R$ and $D$:
\begin{eqnarray}
S(t) = -\beta  I_c(t) - \beta \ell J_c(t), \label{S} \\
R(t) =  \gamma_1   I_c(t) + \gamma_2  J_c(t), \label{R} \\
D(t) =  \delta  I_c(t) + \delta J_c(t). \label{D}
\end{eqnarray}

If we work with the diagnosed recovered and the diagnosed dead,
we get
\begin{eqnarray}
R_J(t) =   \gamma_2  J_c(t), \label{R_J} \\
D_J(t) =   \delta J_c(t). \label{D_J}
\end{eqnarray}

The formulas given here set $t_{\rm in}=0$. To use them with initial
data at a generic $t_{\rm in}$ we just replace $t$ by $t+t_{\rm in}$
in the formulas obtained here.

\vskip 5mm

{\bf Acknowledgements.}
This research has been partially supported by the FEDER /Ministerio de Ciencia, 
Innovaci\'on y Universidades - Agencia Estatal de Investigaci\'on grant No. 
MTM2017-84446-C2-1-R and ENS Paris Saclay program for student interships
abroad.  A. Carpio thanks G. Stadler for nice discussions.

\end{document}